\documentclass{aa} 

\usepackage{soul}
\usepackage{graphicx}
\usepackage{color}
\usepackage{txfonts}
\usepackage{natbib}
\usepackage[squaren,Gray]{SIunits}

\usepackage{bbold}
\usepackage{subcaption}
\usepackage{mwe}

\newcommand{\sink}{\mathrm{sink}}

\newcommand\nmhd{$\textsc{NMHD-F01}$}
\newcommand\imhd{$\textsc{IMHD-F01}$}

\newcommand\nmhds{$\textsc{NMHD-F01}$ }
\newcommand\imhds{$\textsc{IMHD-F01}$ }

\newcommand\nmhdw{$\textsc{NMHD-F01-mu50}$}
\newcommand\nmhdws{$\textsc{NMHD-F01-mu50}$ }

\newcommand\nmhdf{$\textsc{NMHD-F05}$}
\newcommand\nmhdfs{$\textsc{NMHD-F05}$ }

\newcommand\lnmhd{$\textsc{NMHD-F01-M500}$}
\newcommand\lnmhdb{$\textsc{NMHD-BARO-M500}$}

\newcommand\lnmhds{$\textsc{NMHD-F01-M500 }$ }
\newcommand\lnmhdbs{$\textsc{NMHD-BARO-M500 }$}

\begin{document}
  \renewcommand{\vec}[1]{\mathbf{#1}}

   \title{Synthetic populations of protoplanetary disks \\ Impact of magnetic fields and radiative transfer}
   \author{U. Lebreuilly$^{1}$
   \and P. Hennebelle$^{1}$
   \and T. Colman$^{1}$
   \and A. Maury$^{1}$
  \and  N. D. Tung$^{1}$
   \and L. Testi$^{2,3}$
   \and R. Klessen$^{4,5}$
   \and S. Molinari$^{6}$
   \and B. Commerçon$^{7}$
   \and M. Gonz{\'a}lez$^{8}$
   \and E. Pacetti$^{6,9}$
   \and A. Somigliana$^{10,11}$
   \and G. Rosotti $^{12,13}$}

   \institute{$^{1}$Universit\'{e} Paris-Saclay, Universit\'{e} Paris Cité, CEA, CNRS, AIM, 91191, Gif-sur-Yvette, France\\ 
      $^{2}$Dipartimento di Fisica e Astronomia ”Augusto Righi” Viale Berti Pichat 6/2, Bologna \\
    $^{3}$INAF, Osservatorio Astrofisico di Arcetri, Largo E. Fermi 5, I-50125, Firenze, Italy \\
       $^{4}$Universität Heidelberg, Zentrum für Astronomie, Institut für theoretische Astrophysik, Albert-Ueberle-Str. 2, D-69120 Heidelberg, Germany\\ 
      $^{5}$Universität Heidelberg, Interdisziplinäres Zentrum für Wissenschaftliches Rechnen, Im Neuenheimer Feld 205, D-69120 Heidelberg, Germany \\
      $^{6}$INAF - Istituto di Astrofisica e Planetologia Spaziali (INAF-IAPS), Via Fosso del Cavaliere 100, I-00133, Roma, Italy \\
         $^{7}$Univ Lyon, Ens de Lyon, Univ Lyon1, CNRS, Centre de Recherche Astrophysique de Lyon UMR5574, F-69007, Lyon, France \\
      $^{8}$Universit\'{e} Paris Cité, Universit\'{e} Paris-Saclay, CEA, CNRS, AIM, F-91191, Gif-sur-Yvette, France\\ 
        $^{9}$Dipartimento di Fisica, Sapienza Università di Roma, Piazzale Aldo Moro 2, I-00185, Roma, Italy\\
         $^{10}$ESO - European Southern Observatory, Karl-Schwarzschild-Strasse 2, D-85748, Garching bei München, Germany \\
        $^{11}$Fakultat für Physik, Ludwig-Maximilians-Universität München, Scheinerstr. 1, 81679 München, Germany\\
          $^{12}$Dipartimento di Fisica, Università degli Studi di Milano, Via Giovanni Celoria, 16, 20133, Milano, Italy \\
           $^{13}$Leiden Observatory, Leiden University, Niels Bohrweg 2, NL-2333 CA Leiden, The Netherlands\\
              \email{ugo.lebreuilly@cea.fr}      
             }

   \date{}

% \abstract{}{}{}{}{} 
% 5 {} token are mandatory
 
  \abstract
  % context 
   {Protostellar disks are the product of angular momentum conservation during the protostellar collapse. Understanding their formation is crucial because they are the birthplace of planets and because their formation is tightly related to star formation. Unfortunately, the initial properties of Class 0 disks and their evolution are still poorly constrained observationally and theoretically. }
{We aim to better understand the mechanisms that set the statistics of disk properties as well as to study their formation in massive protostellar clumps. We also want to provide the community with synthetic disk populations to better interpret young disk observations. }
     {We use the \texttt{ramses} code to model star and disk formation in massive protostellar clumps with  magnetohydrodynamics including the effect of ambipolar diffusion and radiative transfer including the stellar radiative feedback. Those simulations, resolved up to the astronomical unit scale, allow to investigate the formation of disk populations.  } 
   {Magnetic fields play a crucial role in disk formation. A weaker initial field leads to larger and massive disks and weakens the stellar radiative feedback by increasing fragmentation. We find that ambipolar diffusion impacts disk and star formation and leads to very different disk magnetic properties. The stellar radiative feedback also have a strong influence, increasing the temperature and reducing fragmentation. Comparing our disk populations with observations reveals that our models with a mass-to-flux ratio of 10 seems to better reproduce observed disk sizes. This also sheds light on a tension between models and observations for the disk masses.} 
   {The clump properties and physical modeling impact disk populations significantly.  The tension between observations and models for disk mass estimates is critical to solve with synthetic observations, in particular for our comprehension of planet formation.} 
   \keywords{Hydrodynamics; Magnetohydrodynamics (MHD); Turbulence;  Protoplanetary disks; star: formation }
   
 \maketitle
\section{Introduction}

Protostellar disks, often referred to as protoplanetary disks, are formed through the conservation of angular momentum during the protostellar collapse. New observational evidences suggest that planets, or at least the gas giants, could form early during the evolution of those disks. The mass content of Class II-III disks indeed seems insufficient to explain observed exoplanetary systems \citep{2018A&A...618L...3M,2020A&A...640A..19T}. In addition, the sub-structures of young $<1~$Myr Class II \citep[e.g., in HL-tau][]{2015ApJ...808L...3A} and even $<0.5~$Myr Class I \citep{2020Natur.586..228S} disks, in particular rings and gaps, could be indications of the presence of already formed giant planets. There are, of course, other theories for the formation of those structures \citep[see the recent review by][]{2022arXiv221013314B}, but the hypothesis of the presence of planets in gaps has recently been strengthened by kinematics evidences \citep{2018ApJ...860L..13P,2019NatAs...3.1109P}. In contrast to older disks, Class 0-I disks could still have enough material to form planets. Unfortunately, the properties of these young disks are yet very poorly constrained. They are deeply embedded in a dense protostellar envelope, dominating the mass of protostellar objects during the whole Class 0 phase and are often spatially unresolved in the wavelength range at which they can be observed \citep{2019A&A...621A..76M,2022ApJ...929...76S}. 

On the theoretical perspective, one must resort to large scale simulations self-consistently forming disk populations. Significant efforts toward this challenging modeling have been made by various teams in the pasts. For instance, \cite{Kuffmeier2017}, and later \cite{Kuffmeier2019}, investigated the impact of the accretion from large giant molecular cloud scales on the properties of disks, but without focusing on the formation of a full disk populations. This was done the first time by \cite{2018MNRAS.475.5618B} who investigated a full disk population forming in a massive protostellar clumps. Subsequently, \cite{2021MNRAS.508.5279E} investigated the impact of metallicity on disk population formation in very similar calculations, initially presented by \cite{2019MNRAS.484.2341B}. They mainly concluded that disk radii were decreasing with a decreasing metallicity. However, both studies did not account for the impact of the magnetic field.

Magnetic fields are, however, ubiquitous in observations of Young Stellar Objects \citep[YSOs, e.g.,][]{2006Sci...313..812G,2009ApJ...707..921R,2018MNRAS.477.2760M}. Observations suggest that they may play a key role in shaping the properties of some key features of the star formation process, such as the development of accretion flows, the disk sizes ans masses and the occurrence of multiple stellar systems \citep{2018MNRAS.477.2760M,2020A&A...644A..47G,2023A&A...669A..90C}. On the theory side, their role  has been extensively investigated in the ideal  \citep{PriceBate2007,2008ApJ...681.1356M,2008A&A...477....9H,2008A&A...477...25H,2012A&A...543A.128J} and non-ideal \citep{2009ApJ...706L..46D,2010A&A...521L..56D,2011ApJ...729...42M,2011ApJ...738..180L,2012A&A...541A..35D,2014prpl.conf..173L,2015ApJ...801..117T,2015ApJ...810L..26T,2016A&A...592A..18M,2016A&A...587A..32M,2018A&A...615A...5V,2019MNRAS.486.2587W,2020MNRAS.492.3375Z,2020A&A...635A..67H,2020ApJ...900..180M,2021MNRAS.505.5142Z,2021A&A...652A..69M,2021A&A...656A..85M} MHD frameworks for isolated collapse calculations of low and high mass cores. The magnetic fields have been proven critical to shape the disk through the regulation of angular momentum and for the launching of protostellar outflows. The importance of the large (clump-scale) magnetic field was also pointed out in the zoom-in simulations of \cite{Kuffmeier2017} and \cite{Kuffmeier2019}. In the context of $50 M_{\odot}$ mass clumps, \cite{2019MNRAS.489.1719W}, investigated the effect of all three non-ideal MHD effects on disk formation and concluded that they mostly impacted the small scales and the magnetic properties of the disks but not their size and mass.

So far, only \cite{Lebreuilly2021} investigated disk formation in the MHD context with ambipolar diffusion for massive clump calculations while systematically resolving the disk scales and concluded that the clump scale magnetic field was indeed playing a major role in setting the initial statistical conditions of the disks. Here we continue this work by expanding the parameter space of the simulation suite with an overall higher numerical resolution. In this paper and its companion paper,  \cite{2023arXiv230905397L}, we investigate in detail the impact of the initial clump conditions, i.e. the magnetic field strength, the treatment of the RT and the protostellar feedback (accretion luminosity, jets) and the clump mass. In this work, we will present six models and investigate the impact of the magnetic field and the RT modeling on the initial conditions of protostellar disks.
This article is decomposed as follows. In Sect.~2, we briefly recall our methods, that are similar to those of \cite{Lebreuilly2021}. In Sect.~3, we present in detail our fiducial model. This model will be our reference for comparison in this series of paper. In Sect.~4, we investigate the impact of the magnetic field and RT treatment on the initial conditions of our disk populations and their evolution. In Sect.~5, we describe the main caveats/prospects of our study and a first comparison with observation are then presented. Finally, we present our conclusions in Sect.~6.

\section{Methods}
\subsection{Dynamical equations}
To accurately describe the relevant physics in the context of star and disk formation, we solve the following dynamical equations
\begin{eqnarray}
\frac{\partial \rho}{\partial t} &+&\nabla \cdot \left[ \rho \vec{v} \right] = \nonumber 0,\\
\frac{\partial \rho \vec{v}}{\partial t}  &+& \nabla \cdot \left[ \rho \vec{v} \vec{v} + (P_{\rm{th}} + \frac{\vec{B}^2}{2}) \mathbb{I} -\vec{B} \vec{B} \right] = - \rho  \vec{\nabla} \phi -\lambda \nabla E_{\rm{r}}, \nonumber \\
\frac{\partial E}{\partial t}  &+&\nabla \cdot \left[\vec{v} (P_{\rm{th}} +E +\frac{\vec{B}^2}{2}) - \vec{B (\vec{B} \cdot \vec{v)}}\right]= - \rho \vec{v} \cdot \vec{\nabla} \phi   \nonumber \\  &+&\Lambda_{\rm{AD}}-\mathbb{P}_{\rm{r}} \nabla{:} \vec{v} -\lambda \vec{v} \nabla \cdot E_{\rm{r}}   \nonumber \\ 
&+& \nabla \cdot \left(\frac{c \lambda}{\rho \kappa_{\rm{R}}} \nabla E_{\rm{r}} \right) +  S_{\star} ,\nonumber \\
\frac{\partial E_{\rm{r}}}{\partial t}  &+&\nabla \cdot \left[\vec{v} E_{\rm{r}} \right] = - \mathbb{P}_{\rm{r}} \nabla{:} \vec{v}  + \nabla \cdot \left(\frac{c \lambda}{\rho \kappa_{\rm{R}}} \nabla E_{\rm{r}} \right) 
\nonumber \\
&+& \kappa_{\rm{P}} \rho c (a_{\rm{R}} T^4- E_{\rm{r}}) +  S_{\star} , \nonumber \\
\frac{\partial \vec{B}}{\partial t}  &-& \nabla \times \left[\vec{v} \times \vec{B} \right] 
\nonumber \\
&+&\nabla \times \frac{\eta_{\rm{AD}}}{|\vec{B}|^2} \left[((\nabla \times \vec{B}) \times \vec{B} )\times \vec{B}\right] = 0, \nonumber \\
\nabla \cdot \vec{B}&=&0, \nonumber \\
\triangle \phi &=& 4 \pi \mathcal{G} \rho,
\end{eqnarray}
where $\rho$ and $\vec{v}$, $E$, $E_r$ and $\vec{B}$ are the gas density and velocity, the total energy, the radiative energy and the magnetic field. We also define the thermal pressure $P_{\rm{th}}$, the gravitational potential $\phi$, the radiative pressure $\mathbb{P}_{\rm{r}}$, the Rosseland $\kappa_{\rm{R}}$ and Planck opacities $\kappa_{\rm{P}}$, the radiative flux limiter $\lambda$ \citep{1978JQSRT..20..541M}, the temperature $T$, the total luminosity source term $S_{\star}$, the ambipolar resistivity $\eta_{\rm{AD}}$ and $\Lambda_{\rm{AD}}$ the heating term due to ambipolar diffusion. Finally, we define the gravitational constant $\mathcal{G}$, the Stefan-Boltzmann constant $a_{\rm{R}}$ and the speed of light $c$.

To solve these equations, we use the adaptive mesh-refinement code (AMR) {\ttfamily ramses} \citep{2002A&A...385..337T,2006A&A...457..371F} with RT in the flux-limited diffusion (FLD) approximation \citep{2011A&A...529A..35C,2014A&A...563A..11C}, non-ideal MHD, and more particularly ambipolar diffusion \citep{2012ApJS..201...24M} and sink particles \citep{2014MNRAS.445.4015B}. More details about the code and modules that we used in this study can be found in the works mentioned above.

\subsection{Initial conditions}

Our clumps are initially uniform spheres of $500-1000 M_{\odot}$, of temperature $T_0=10~$K and with an initial radius given by the thermal-to-gravitational energy ratio $\alpha$ such as 
\begin{eqnarray}
\alpha \equiv \frac{5}{2} \frac{R_0 k_{\rm{B}} T_{0}}{\mathcal{G} M_0\mu_{\rm{g}}m_{\rm{H}}},
\end{eqnarray}
  $k_{\rm{B}}$ being the Boltzmann constant, $m_{\rm{H}}$ the Hydrogen atom mass and $\mu_{\rm{g}}=2.31$ the mean molecular weight. Owing to our choice for the values of $\alpha$, all our clumps have the same initial radius of $\sim 0.38~$pc. The box size $L_{\mathrm{box}}$ is chosen to be four times larger, i.e., $L_{\mathrm{box}}=1.53$~pc. Outside of the clump the density is divided by 100.
We set an initial turbulent velocity at Mach 7 with a Kolmogorov powerspectrum of $k^{-11/3}$ and random phases to mimic the molecular cloud turbulence. We point out that, as explained in \cite{LeeandHennebelle2018}, the initial choice of spectrum for the turbulence has little impact on the results because the initial conditions are quickly forgotten has the collapse proceeds. Better ways to model the turbulence would require to start the simulation from even larger (kpc) scales, which is clearly way beyond the scope of the present work.

The magnetic field strength is initialised according to the  mass-to-flux over critical-mass-to-flux ratio $\mu$ such as
\begin{equation}
    \mu = \left(\frac{M_0}{\phi}\right)/\left(\frac{M}{\phi}\right)_c,
\end{equation}
with $\left(\frac{M}{\phi}\right)_c = \frac{0.53}{\pi}\sqrt{5/\mathcal{G}}$ \citep{1976ApJ...210..326M}.

\subsection{Sink particles}

Sink particles, following the implementation detailed in \citep{2014MNRAS.445.4015B}, are employed to mimic the behaviour of stars in our models. They are formed when the local density reaches the density threshold $n_{\mathrm{thre}}= 10 ^{13}~ \centi\meter^{-3}$. This value is chosen in accordance with the analytical estimate of \cite{2020A&A...635A..67H}. Once we form a sink, it automatically accretes, at each timestep, the material with a density above the threshold and within the sink accretion radius $4 \Delta x$. 

\subsection{Radiative transfer modelling}
We examine two possible ways of modelling the RT. For all models, except one (\lnmhdb), we include the RT in the FLD approximation, using the solver of \cite{2011A&A...529A..35C,2014A&A...563A..11C}. 

In this approach we consider sinks/stars as sources of luminosity, defined as the sum of the intrinsic and accretion luminosity. The former is computed using the evolutionary tracks of \cite{2013ApJ...772...61K}, while the latter is defined as 
\begin{eqnarray}
L_{\rm{acc}} = f_{\rm{acc}} \frac{\mathcal{G} M_{\sink} \dot{M}_{\sink}}{R_{\star}},
\end{eqnarray}
where $R_{\star}$ is the star radius, also extracted from the tracks of \cite{2013ApJ...772...61K}, $M_{\sink}$ and $\dot{M}_{\sink}$ are its mass and mass accretion rate, and $0<f_{\rm{acc}}<1$ is a dimensionless coefficient. $f_{\rm{acc}}$ corresponds to the amount of gravitational energy converted into radiation, in this work, we explore two values for $f_{\rm{acc}}$ equal to 0.1 and 0.5. We refer to \cite{Lebreuilly2021} for an explanation on how the luminosity source terms are implemented in the code.

In the second approach (run \lnmhdb) we do not use the FLD approximation but instead assume a barotropic equation of state (EOS) to compute the temperature. This is of course an over-simplification, but it is interesting for two main reasons. First, barotropic EOS models are have lower temperatures than FLD calculations \citep{2010A&A...510L...3C}, they allow to investigate the effect of temperature on the disk formation and evolution. Second, because they are simpler than a full radiative tranfer modeling, these EOS are still widely used by the community. 

For our barotropic EOS calculation, we assume \cite[as in][]{2016A&A...592A..18M}
\begin{equation}
    T =  T_{0} \sqrt{1+\left(\frac{n}{n_1}\right)^{0.8}}\left[1+\left(\frac{n}{n_2}\right)\right]^{-0.3},
\end{equation}
with $n$ the gas number density, $n_1=10^{11}~ \centi\meter^{-3}$, $n_2=10^{16}~ \centi\meter^{-3}$, and $T_0= 10~\kelvin$.
\subsection{Refinement criterion}

 We use the AMR grid of {\ttfamily ramses} which allows us to locally refine the grid according to the local Jeans length. 
 More specifically, we use a modified Jeans length such as 
 \begin{equation}
    \tilde{\lambda_{\mathrm{J}}}= \left\{\begin{array}{ll}
    \lambda_{\mathrm{J}}  & \mathrm{~if ~} n< 10^9 ~\centi\meter^{-3},\\
    \mathrm{min} (\lambda_{\mathrm{J}},\lambda_{\mathrm{J}} (T_{\mathrm{iso}})) & \mathrm{~otherwise ~}
    \end{array}
\right.
\end{equation}
This modification is convenient for studying disk formation especially in the presence of feedback since it is independent from the temperature at $T>T_{\mathrm{iso}}\equiv 300~$K in the dense and heated regions.  
 In all our models, we impose 10 points per modified Jeans lengths within each cell to prevent artificial fragmentation \citep{1997ApJ...489L.179T}. The cell size is computed as a function of the refinement level $\ell$ as 
\begin{eqnarray}
\Delta x = \frac{L_{\mathrm{box}}}{2^\ell}.
\end{eqnarray}

Our resolution always ranges from $\sim 2460~$au ($\sim 0.012~$pc) in the coarsest cells of the simulation down to $\sim 1.2~$au ($\sim 5.8 \times 10^{-6}~$pc) in the fine cells.

\subsection{Disk selection}
\label{sec:diskextract}
The disks analyzed in the present study were selected using the same method as in \cite{Lebreuilly2021}, but we slightly modified the pre-selection criteria of \cite{2012A&A...543A.128J}. As a reminder,  the Joos criterion are 
\begin{itemize}
    \item $n>10^{9}~\centi\meter^{-3}$, where $n$ is the number density;
    \item $v_{\phi}>2 v_{r}$,  $v_{\phi}>2 v_{z}$, where $v_{r}$, $v_{z}$ and $v_{\phi}$ are the radial, vertical and azimuthal velocities, the rotation axis being the direction of the angular momentum at the sink vicinity;
    \item $1/2 \rho v_{\phi}^2>2 P_{\mathrm{th}}$, where $\rho$ is the gas density and $P_{\mathrm{th}}$ is the thermal pressure.
\end{itemize}
In this work, we consider only the two first criteria. We have found that the last energy criterion arbitrarily removes the inner hot regions of the disks. Removing this criterion also allows a better comparison of models with a different accretion luminosity efficiency (and hence different temperatures).

Once all the disks of a model are selected, we analyze various of their internal properties (their radius, mass, temperature etc.). 
 For any quantity A, we compute volume average such as 
\begin{eqnarray}
    \left<A\right> = \frac{\sum_j A_j \Delta x_j^3}{\sum_j \Delta x_j^3},
\end{eqnarray}
where $j$ refers to all the disk cells of size $\Delta x_j$. The treatment of the temperature is slightly different, as we select only mid-plane cells to compute its averaged value. This allows a better estimate of the temperature in the hot regions of the disk. In the remaining of the manuscript, we drop the $\left<\right>$ notation for averages as no confusion with the local value is possible. 
In addition, we estimate the disk radius as the median of the maximal extent in 50 equal-size azimuthal slices and the disk mass as the sum of the mass of every disk cell. 
\subsection{List of models}

\begin{table*}
  \caption{summary of the different simulations. From the left to the right: model name, initial clump mass, thermal-to-gravitational energy ratio $\alpha$, mass-to-flux ratio $\mu$ ($\star$ means ideal MHD), accretion luminosity efficiency $f_{\mathrm{acc}}$ (if applicable), RT modeling, final median sink mass and final SFE and corresponding time $t_{\mathrm{end}}$.}      
\label{tab:modelsdisk}      
\centering          
\begin{tabular}{c c c c c c c c c c }     
\hline\hline       
                   Model name &  Mass [$M_{\odot}$] &$\alpha$ & $\mu$ &  $f_{\mathrm{acc}}$ & RT  & $N_{\mathrm{sinks}}$ & Median sink mass [$M_{\odot}$] &Final SFE  & $t_{\mathrm{end}}$\\ 
\hline                    
\nmhd &  1000  & 0.008 & 10  & 0.1 & FLD & 88 & 0.9& 0.15 & 116.6\\
\nmhdf &  1000  & 0.008  & 10 & 0.5 & FLD & 63  & 1 &0.15 & 117.5\\
\imhd &  1000   & 0.008 &10$\star$ & 0.1& FLD & 61 & 1& 0.15 & 118.5 \\
\nmhdw&  1000  & 0.008  & 50 & 0.1& FLD & 138 & 0.37&0.15 & 110.4\\
\lnmhd &  500  & 0.016 & 10  & 0.1 & FLD & 106 & 0.37&0.11 &162\\
\lnmhdb &  500  & 0.016 & 10  & - & Barotropic & 147 & 0.15 &0.11 & 157\\

\hline \hline
\end{tabular}
\end{table*}
Our full list of models computed for this work is presented in Tab.~\ref{tab:modelsdisk}. From left to right the table shows the model name, the initial clump mass, the thermal-to-gravitational energy ratio $\alpha$, the mass-to-flux ratio $\mu$ ($\star$ means ideal MHD), the accretion luminosity efficiency $f_{\mathrm{acc}}$ (if applicable), the choice of RT modeling, the final median sink mass and the final SFE and corresponding time $t_{\mathrm{end}}$. 
\section{Presentation of the fiducial model} 

Our reference model \nmhds is a $1000~ M_{\odot}$ clump with $\mu=10$, $f_{\mathrm{acc}}=0.1$ and a Mach number of 7. This run is essentially the same as the fiducial calculation of \cite{Lebreuilly2021}, but with our new improved refinement criterion.

\subsection{Large scales and star formation}

\begin{figure*}[h!]
  \centering
 \includegraphics[width=
          \textwidth]{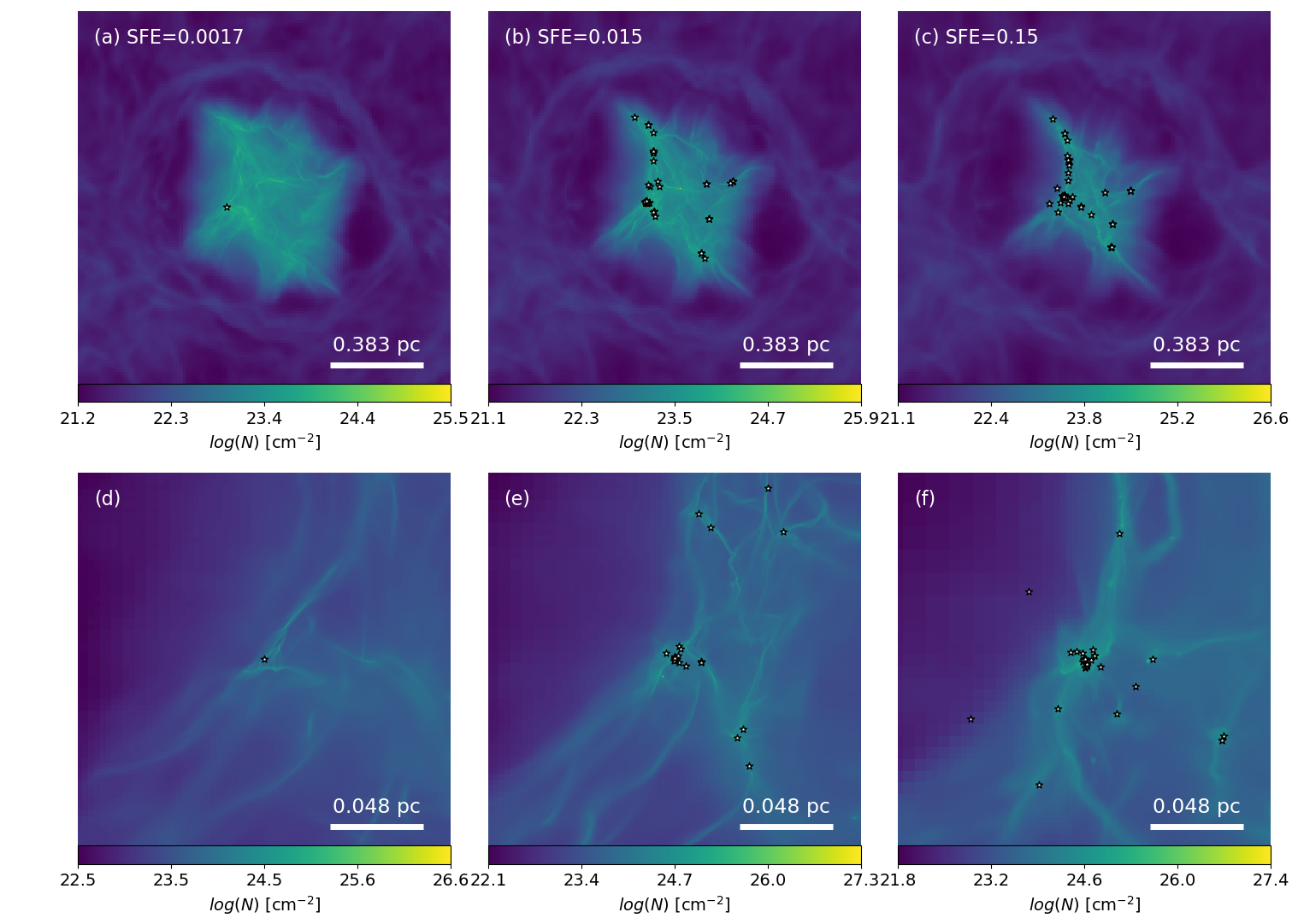}
      \caption{Evolution of the clump of run \nmhds at various times (SFE=0.0017, 0.015 and 0.15). (a,b,c) Full column density maps in the (x-y) plane. (d,e,f) Same but centered around the sink 1 (located in the hub) and with an extent of $12.5\%$ of the box scale. Sinks are represented by the star symbols.} 
            \label{fig:column_fid}
\end{figure*}

We begin the description of our fiducial run \nmhds by briefly presenting its evolution at the global scale. The panels (a), (b) and (c) of Fig. \ref{fig:column_fid} show column density snapshots at various evolutionary stages for this model. The sinks are represented by the star markers. 

As expected, the gravoturbulent motions lead to the formation of a network of highly non-homogeneous filament structures along which star formation mainly occurs \citep[as in other similar studies, e.g.,][]{LeeandHennebelle2018,2018MNRAS.475.5618B,Lebreuilly2021,Grudic2021,turbsphere}. It is also very clear that the stars do not form in isolation here. Star formation is in fact more  concentrated around one compact hub in the bottom-left part of the clump. This effect, which is most certainly a consequence of the global collapse, off-centered because of the turbulence, was also observed in the models of \cite{Lebreuilly2021} and \cite{Hennebelleetal2022}. The global collapse of the clump is quite noticeable and non-isotropic, as expected in the presence of turbulence, which explains the presence of a main star-forming filament. This filament, connected with the previously mentioned hub, is the second most active site of star formation of the clump.

We now focus on the main hub evolution. The panels (d), (e) and (f) of Fig. \ref{fig:column_fid} show the column density of \nmhds at the same SFE as the top panels but centered around the hub and at a smaller scale ($12.5\%$ of the box). Even at those scales, stars are clearly formed in filamentary structures which are connected to the larger scale network seen in the top panels of Fig.~\ref{fig:column_fid}. These filaments are similar to the bridge structures that were observed in \cite{Kuffmeier2019}. They connect sinks with their neighbours and represent a shared reservoir of mass. They are relatively quiescent and typically survive a few $\simeq 10~$kyr. Quite clearly, a compact and highly interacting protostellar cluster is formed at the center of the hub. We point out that, although sinks can get quite close to each others in this hub, we chose to never merge the sinks in our models since we are not resolving the stellar radii scales. This hub is a favoured place to form massive stars in the clump. In fact, the most massive stars formed in the model are part of this cluster. 

Between a SFE = 0.015 ($t=97.2~$kyr) and SFE = 0.15 ($t=116.6~$kyr), we observe a clear thickening of the filaments due to the radiative feedback of stars that heat-up the gas and therefore increase its thermal support over time. This increase of thermal support significantly reduces fragmentation and sink formation which essentially halts after a very efficient early phase \citep{Hennebelleetal2022}. Over the course of the simulation, integrated up to SFE=0.15, about 90 sinks are formed, half of which are either single star or primaries  \citep[according to the simplified definition of][]{Lebreuilly2021}.

We point out that this model has formed less stars than its lower resolution counterpart  \citep[the nmhd model of ][]{Lebreuilly2021}. Very interestingly, the overall higher resolution of \nmhds allows the formation of one massive $\sim 15 M_{\odot}$ star over the course of the simulation. This is more massive by a factor of a few than the ones obtained in lower resolution runs \citep{Hennebelleetal2022}. We stress a clear correlation between sink masses and the mass of their surrounding envelope at a $1000~$au scale which indicates that the more massive star form in the more massive environment \citep[see also,][]{2000ApJS..128..287K,2020MNRAS.492.4727C}.  

For more extended dedicated descriptions (temperature, magnetic field and stellar mass spectrum) of the clump scales and star formation in very similar calculations, we refer the reader to \cite{Hennebelleetal2022} and references therein.

\subsection{Disks and small scales}

\begin{figure*}[h!]
\centering
 \includegraphics[width=
          0.85\textwidth]{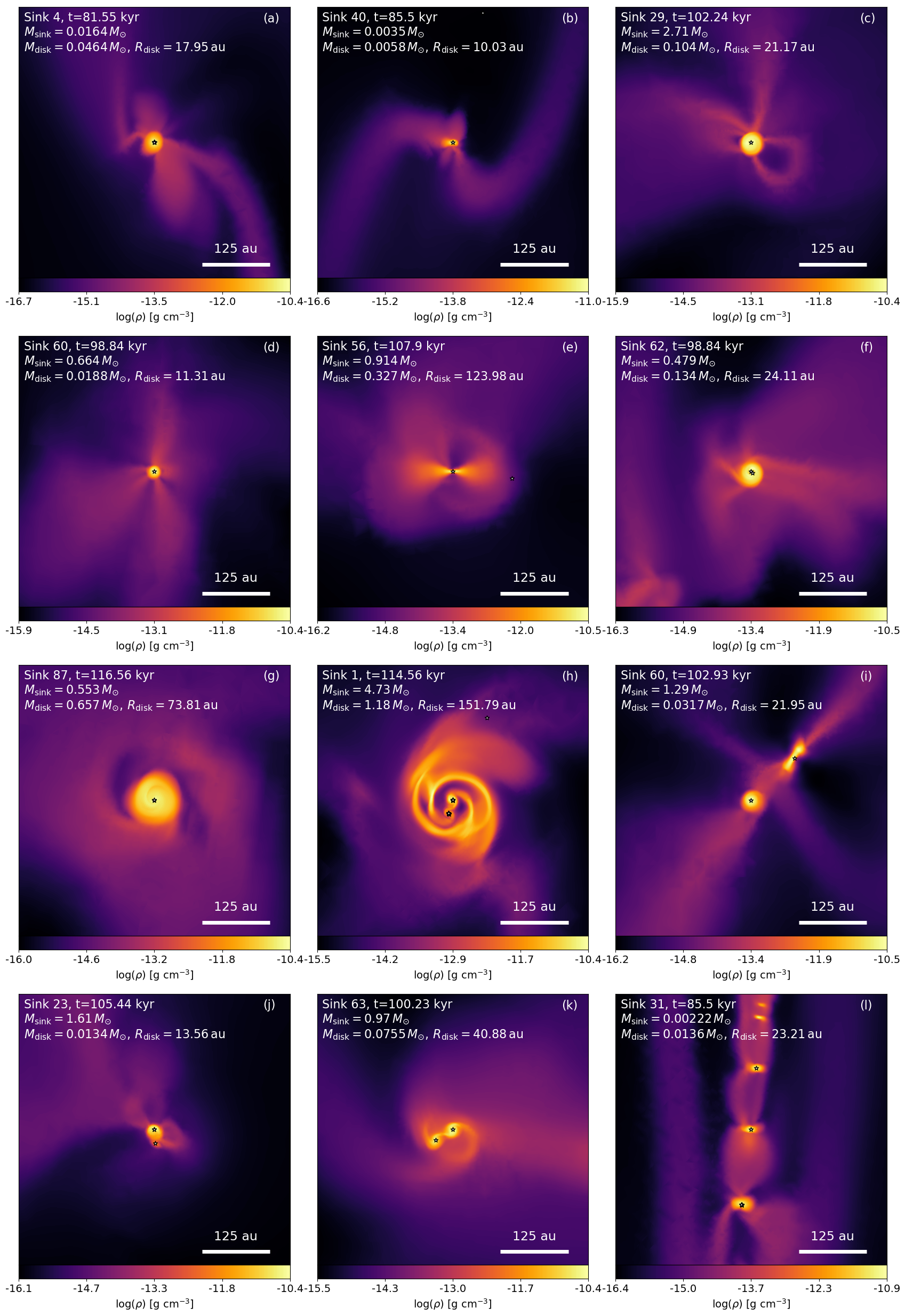}
     \caption{collection of disks from run \nmhd; edge-on or mid-plane density slices. In addition to the density, for each disk we display the sink index, the time of the corresponding snapshot, the mass of the sink and of the disk as well as the disk radius. Some disks, displayed at different time can appear in several panels.} 
    \label{fig:collec_fid}
\end{figure*}

In the following, we describe further the formation of structures, their properties and their evolution at disk scales. 

There is a clear variety of disks and a wide diversity of commonly observed small scale features in the \nmhds model and, more generally, in all of our models. We describe here the typical appearance and structures of our disks. As a support for that description we show, in Fig.~\ref{fig:collec_fid}, edge-on or mid-plane density slices of 12 of our fiducial model disks at various times. 
\begin{itemize}
\item{Compact disks:} first of all, we observe many compact disks (panels a, b, c, d, f, k, i, j, k and l). They, in fact, represent the majority of the disks in the model as we show later in this section (half of the disks are smaller than $\sim 28$~au at birth). As explained in \cite{Lebreuilly2021}, they are a clear consequence of the regulation of the angular momentum by the magnetic braking during the protostellar collapse. It is worth mentioning  that disks are indeed expected, from observations, to be compact at the Class 0 and early Class I stage \citep{2019A&A...621A..76M}. The frequent occurrence of these $< 50$~au disks is compatible with the self-regulated scenario of \cite{2016ApJ...830L...8H}. In this scenario, it is expected that the interplay between magnetic braking and ambipolar diffusion mostly leads to the formation of compact disks.

\item{Sub-structures:}
spirals/arcs (panel g,h,k) are also often observed, particularly in the presence of multiple systems or/and when the disk is gravitationally unstable and fragmenting. The latter effect is however restricted to the most massive and young disks (while they are still cold) as it is quite efficiently suppressed by the stellar feedback.
Noticeably, ring structures are not present in our models: often attributed to planets-disk interactions, they could in principle form from MHD instabilities in the disks \citep[see the recent review by][]{2022arXiv220309821L}. The fact that we do not observe then could come from two reasons, namely that either our resolution is still insufficient for them to occur in our disks or the early conditions in the disks (hot disks, with a massive turbulent envelope) are not favorable for rings to form. In general the disk sub-structures are quite faint unless they originate from multiplicity in the disk (e.g panel h). This is most likely a consequence of the thermal support due to radiation that stabilises the disk structure.

\item{Magnetised flows:} a common consequence of the magnetic field interplay with the gas at the disk scale is the triggering of interchange instabilities in some cases (as revealed by the prominent loop seen in panel c). This instability, that can transport momentum away from the disks \citep{2012ApJ...757...77K}, was also observed in the rezoomed models of \cite{Kuffmeier2017} in ideal MHD. Our study confirm that they are not always suppressed by the diffusive effect of ambipolar diffusion. Quite noticeably, and as was also noted in \cite{Lebreuilly2021}, magneto-centrifugal outflows and jets are however absent in the models. This is most likely due to a suppression by the turbulence for the former \citep[in accordance to][]{2021A&A...656A..85M} and a lack of resolution in the inner regions of the disks for the latter. In Paper II, we will investigate the impact of protostellar jets on the disk properties using the sub-grid modeling of \cite{2022A&A...663A...6V}.

\item{Non-axisymmetric envelopes:} the protostellar envelope around the disks are still massive by the end of the computation. Generally speaking, they are very diverse in structure and never spherically symmetrical, as already pointed out by \cite{Kuffmeier2017}. From as far as a few $\sim 1000~$au scale up to the disk scales, accretion streamers are very common around disks in all the simulations. Those channels for high density material accretion are typically (but not exclusively) connected to the disk mid-plane (e.g., panels a and b).

\item{Flybys:}
close interactions between stars (not orbiting each other) or between a star and dense clumpy gas are common during the clump evolution \citep[][and references therein]{2022arXiv220709752C}. Close flybys of two (or more) disks (panel i, j and k) are happening several times in the model. In our simulations, they quite systematically lead to disk mergers, probably through the bridge structures \citep{Kuffmeier2019}. Flybys have been shown to be able to truncate disks and trigger spiral formation in idealized calculations \citep{2019MNRAS.483.4114C}. It is however difficult to establish their role in complex clump simulations that include many potential mechanisms to generate structures in the disks.

\item{Disks in columns:} a very peculiar structure, with several occurrences at the early stages of the calculations (in all the models) only, is the formation of several disks in a single filament/column (panel l). It is the consequence of the fragmentation of the clump at the filament scale. This is only possible when the latter are still cold. Indeed, this behaviour is not observed later when the stellar feedback heats up the gas and the thermal support precludes further fragmentation. The very short timescale and early disappearance (in less than 10 kyr after the first sink is formed) of these structures probably explains why they have not been observed.

\end{itemize}

\section{Disk populations}

In the following, we discuss the properties of the disk populations for all of our models. They are extracted for each model as explained in section~\ref{sec:diskextract}. 
\subsection{Fiducial model}

\begin{table}
      \caption{mean, median (med.) and standard deviation (Stdev.) of the disk properties for \nmhds at birth time and at age of 10 and 20 kyr. (-N) stands for $\times 10^{-N}$.}      
\label{tab:propsdisknmhd}      
\centering          
\begin{tabular}{c c c c }     

$R_{\mathrm{disk}}$ [au] & Mean & Med. & Stdev. \\
\hline

At birth & 47.1 & 28.2 & 47.25 \\
10 kyr & 50.7 & 47.2 & 18.15 \\
20 kyr& 59 & 47.2 & 32.6 \\
\hline  \\
$M_{\mathrm{disk}}$ [$M_{\odot}$] & &  &  \\
\hline  

At birth &  8.7(-2) & 3.1(-2) & 0.14\\
10 kyr &9(-2) &  7(-2)& 0.1\\
20 kyr& 0.1 & 8(-2) &  9(-2)\\
\hline \\
Disk-to-stellar mass ratio &  &\\
\hline  

At birth & 4.15  & 1 & 12\\
10 kyr & 3.9 & 0.1 & 15 \\
20 kyr& 2.7 & 4(-2)  &10.4  \\
\hline \\

$T_{\mathrm{mid,disk}}$ [K] &  &\\
\hline  

At birth & 139   &100 & 125\\
10 kyr & 260 &235  & 153 \\
20 kyr& 270&  317 &  226\\
\hline \\

$|B|$ [G] &  &\\
\hline  

At birth & 4(-2)  & 3(-2) &3.7(-2) \\
10 kyr & 0.1 & 7(-2) & 8(-2) \\
20 kyr& 0.14 &0.1   & 0.11 \\
\hline \\
$\beta$ &  &\\
\hline  

At birth & 76  &39 &123 \\
10 kyr & 21 & 12 & 30 \\
20 kyr& 10&   8& 12 \\
\hline \hline
\end{tabular}
\end{table}
\begin{figure*}[h!]
  \centering
 \includegraphics[width=
          \textwidth]{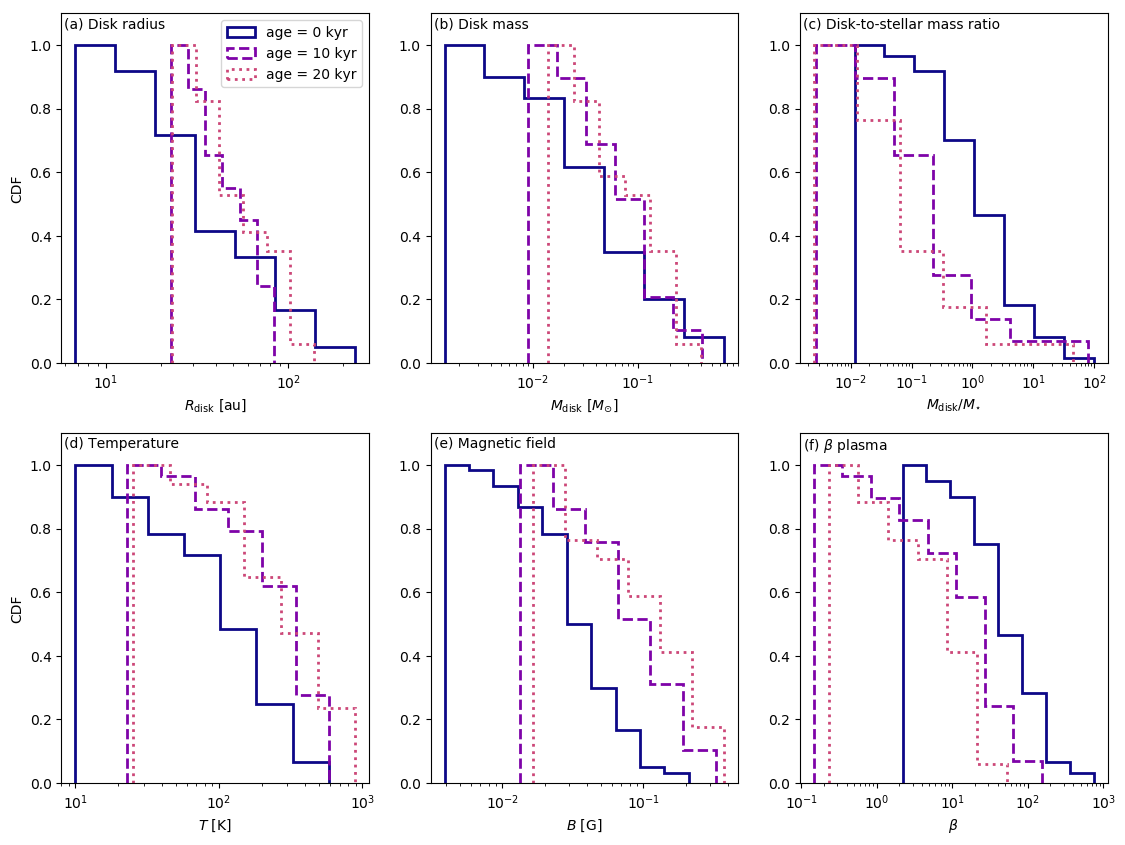}
\caption{CDF of the disks at birth time, but also 10 and 20 kyr after for the \nmhds model.} 
        \label{fig:init_distri_no_cumul}
\end{figure*}

The disk population of the \nmhds model are described in detail in this section.  

In Fig.~\ref{fig:init_distri_no_cumul} we show the cumulative density function (CDF) of several key disk properties for run \nmhds at the closest output from their birth time, but also 10 and 20 kyr after. We show the CDF of the radius (a), the mass (b), the ratio between the disk mass and the stellar mass (hereafter disk-to-stellar mass ratio, panel c), the mid-plane temperature (d), the magnetic field (e) and finally, the plasma beta $\beta \equiv 8 \pi P_{\mathrm{th}} /|B|^2$ (i.e., the ratio between the thermal pressure and the magnetic pressure, panel f). Before describing those quantities individually, we point out that the disk populations are not strongly varying over time in the statistical sense except during the initial 10 kyr (hereafter the disk build-up phase). This does not mean however that disks are not individually evolving. As shown above, they are indeed non-linearly affected by interactions with the clump and other disks. In addition from the histograms, we show in Tab.~\ref{tab:propsdisknmhd} the mean, median (med.) and standard deviation (Stdev.) of these disk properties for \nmhds at birth time, at age of 10 and also 20 kyr and more.

It is worth mentioning that about $\sim 25$ disks are steadily detected in our fiducial model. This number although lower than the findings of \cite{Lebreuilly2021}, is consistent with the reduced fragmentation with our improved refinement criterion. The overall disk-to-star number ratio is higher than in this previous study as we now have $\sim 70\%$ of the systems hosting a disk.
\subsubsection{Sizes and masses}
\begin{figure}
             \centering
       \includegraphics[width=
         0.45\textwidth]{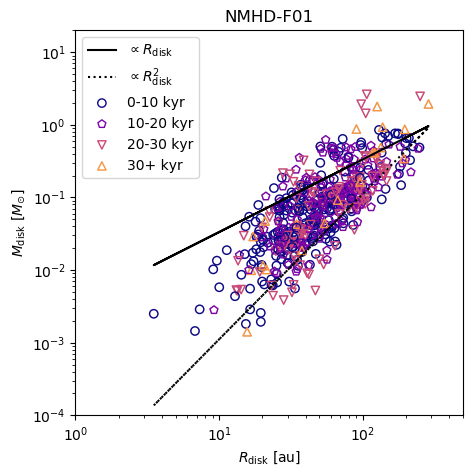}
      \caption{ disk radius vs disk mass for the \nmhd model. Each disk is displayed once per kyr and the different markers/colors represent the various evolutionary stages for the disks.} 
              \label{fig:correlation_radius}
\end{figure}

\begin{figure}
  \centering
         \centering
       \includegraphics[width=
          0.45\textwidth]{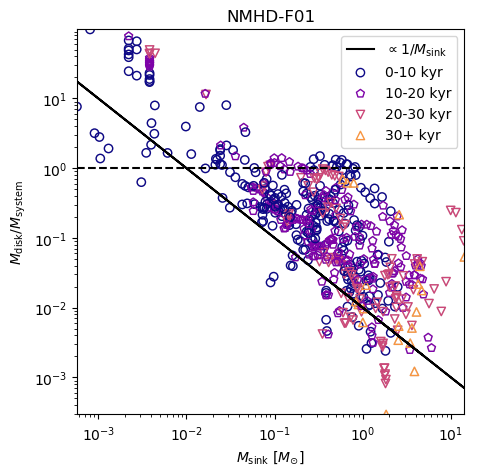}
      \caption{ same as Fig.\ref{fig:correlation_radius} but for the disk-to-stellar mass ratio vs the primary mass. The horizontal dashed line represents a mass ratio of 1.} 
              \label{fig:correlation_disk_to_stell}
\end{figure}

\paragraph{Size} We first focus on the disk radii. This quantity is very interesting because it is perhaps the most reliable observable at all evolutionary stages. We can see, in the radius CDF, that the disks of \nmhds are typically compact,  half of them being smaller than $30~$au. This is in good agreement with the observations \citep{2019A&A...621A..76M,2022ApJ...929...76S} and is slightly lower than what we found in \cite{Lebreuilly2021}. This was however expected, as we speculated that a higher resolution was needed to model the smaller disks. An interesting aspect of the evolution of the disk radius is that the radius of the smallest disk in the sample shifts from less than $10~$au at birth to about $25~$au for older disks. At the same time, the median value shifts from $\sim 28~$au to $\sim 47$~au. However, from 10 to 20 kyr, the PDF does not evolve significantly. To summarize, there is an initial build-up phase in the first 10 kyr of the disks lifetime during which they evolve a lot from compact to  more extended disks as they (and their host star) accrete material from the envelope \citep{2020A&A...635A..67H}. Since this timescale is short, it would be unfortunately almost impossible, for statistical reasons, to observe these disks. After this phase, the disk size does not change much with time, except in case of external perturbations (e.g., flybys/mergers) or when the system is a multiple. 

\paragraph{Mass}  It is of great importance to quantify accurately the disk mass since the mass content of the disks gives valuable information about the budget that is available for planet formation. An useful quantity to keep in mind is the Minimum Solar Mass Nebula \cite[][MSMN]{1981PThPS..70...35H}. This mass, of the order of $0.01 M_{\odot}$ and revised downward in more recent studies \citep[e.g.,][]{2007ApJ...671..878D}, gives the minimal content needed to form solar-like systems. In addition, the disk-to-stellar mass ratio provides an insight on the importance of self gravity in the disk dynamics.  Similarly to the disk radius distribution, there is a clear evolution of the disk mass and disk-to-stellar mass distribution during the build-up of the disks, i.e., over $\sim 10~$kyr. During the early stages of their evolution, a significant fraction of mass in the system still belongs to the disk component which is comparable, if not larger, to the mass of the stellar component.  This is the stage during which the most massive disks can be gravitationally unstable, which could have consequences for early planet formation. After the bulk of the disk mass has been accreted by the protostar, i.e., after a few kyr, the disk typically represents between 1 and 10 $\%$ of the system mass. After this main build-up event the disk masses are still typically larger than their initial value as the disk gets new material from the envelope.  At this stage, half of the disks have masses between $0.01 M_{\odot}$ and slightly less than $0.1 M_{\odot}$ and the other half can reach masses up to $\sim 0.3 M_{\odot}$ which is still more than enough for planet formation according to the MSMN criterion. This value is, indeed, close to, if not below, the low disk mass limit of \nmhds after the build-up phase, the mass of the disks in \nmhds is most likely sufficient to form planetary systems similar to the solar system. Noteworthy, the disk masses that we report are in good agreement with those of the hydrodynamical model of \cite{2018MNRAS.475.5618B}. Noticeably, the disk mass quite clearly correlates with the disk radius. This was of course expected as large volumes with the same typical density contain more mass.  In  Fig.\ref{fig:correlation_radius}, we show this correlation for the \nmhd model. Each disk is  displayed every 1 kyr and the various markers represent the different evolutionary stages of the disks (from birth up to 40 kyr for the older disks). The correlation between disk mass and disk radius is typically in between $\propto R_{\mathrm{disk}}$ (plain line) and $\propto R_{\mathrm{disk}}^2$ (dotted line), and closer to the latter, which is expected for a perfectly symmetric disk. Figure~\ref{fig:correlation_disk_to_stell} shows analogous  information to Fig.~\ref{fig:correlation_radius} for the disk-to-stellar mass ratio vs the stellar mass. A clear correlation between the disk-to-stellar mass ratio and the sink mass is observed. It is slightly steeper, albeit close, to an inversely linear relation. We point out that this correlation also means that the disk mass is only weakly dependent on the stellar mass. In addition, as can be seen, the disks are more massive than the stellar component almost exclusively in the presence of low mass (below $0.1 M_{\odot}$) and young ($<10~$kyr) systems. This is at odds with the previous study of \cite{2018MNRAS.475.5618B} who has found a linear relationship between the disk mass and the stellar mass. We point out that, as discussed extensively in \cite{2020A&A...635A..67H} (see also Sect.~\ref{sec:caveats}), the disk mass depends on the recipe used for the sink particles. This might explain why we do not find the same correlation as \cite{2018MNRAS.475.5618B}. 

\subsubsection{Temperatures}

\begin{figure}
         \centering
       \includegraphics[width=
          0.45\textwidth]{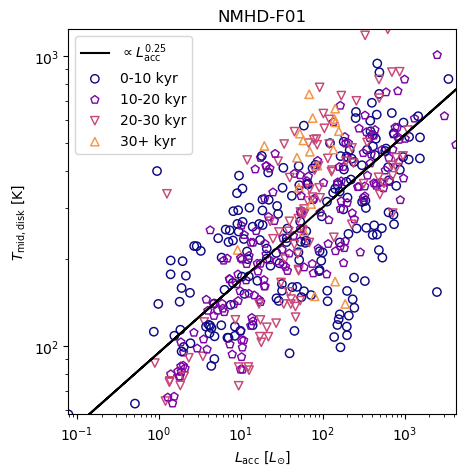}
      \caption{same as Fig.\ref{fig:correlation_radius} but for the mid plane temperature} 
              \label{fig:correlation_T}
\end{figure}

We now turn our attention to the mid-plane temperature distribution. The disks of \nmhds are typically warm, with a median temperature of about $\sim 100~$K during the build up phase and about $\sim 300~$K at later stages. As shown in \cite{2020ApJ...904..194H} and later in \cite{Lee2021}, the temperature at the vicinity of a star can be controlled by its luminosity when it is sufficiently strong. In this context, we expect the disk temperature to scale as $\propto L_{\mathrm{acc}}^{1/4}$. The information that we show in Fig.~\ref{fig:correlation_T} is analogous to Fig.~\ref{fig:correlation_radius}, but for the correlation between disk temperature and their sink accretion luminosity. Those two quantities clearly have a correlation that is close to $T \propto L_{\mathrm{acc}}^{1/4}$ (solid black line). We thus conclude that the accretion luminosity is indeed the dominant factor controlling the disk temperature in the model. More generally, we have confirmed this behavior for all the models, except of course \lnmhdb, for which the temperature is prescribed by the barotropic EOS. 

It is worth mentioning that, despite the weak correlation between the temperature and the accretion luminosity, it leads to a broad distribution of disk mid-plane temperatures (ranging from approximately 60 K to about 1000 K) since the accretion luminosity varies over four orders of magnitude. We have verified that the higher accretion rates and hence accretion luminosity correspond to the most massive stars of the system. As a consequence, the hotter disks are also those around the more massive stars in the model.

\subsubsection{Magnetic fields}

\begin{figure}
  \centering
       \includegraphics[width=
         0.45\textwidth]{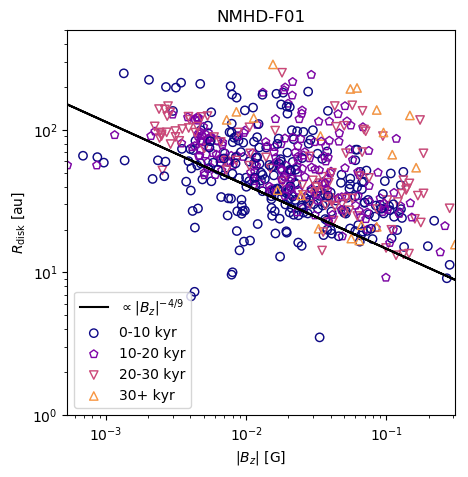}
      \caption{same as Fig.\ref{fig:correlation_radius} for the disk radius vs the absolute value of vertical magnetic field.} 
              \label{fig:correlation_B}
\end{figure}

Finally, we bring our attention to the magnetic field strength and topology in the disks. First of all, half of the disks have a magnetic field below $\sim 0.03$~G during the build-up phase and below $\sim 0.1$~G after. We observe a slight increase of the magnetic field strength during the build-up phase which is probably due to an amplification through the still significant infalling motions that bend the field lines. The distribution of magnetic field is not varying significantly afterward. We also point out that none of the disks has a magnetic field larger than $\sim 0.3~$G. These are the two main consequences of ambipolar diffusion.  The imperfect coupling between the neutrals and the ions indeed leads to a diffusion of the magnetic field at high density where it is not dragged anymore by the gas motions \citep[see][for similar observations in low and high-mass isolated collapse calculations]{2016A&A...587A..32M,2021A&A...652A..69M,2021A&A...656A..85M,2022A&A...658A..52C}. A consequence of the low disk magnetisation is the negligible role played by the magnetic pressure inside the disk. The thermal pressure, enhanced by the stellar feedback from the accretion luminosity, is the main source of support (beside rotation) in the disk. We find that the magnetic pressure is typically about one order of magnitude lower than the thermal pressure and that a majority of disks have $1<\beta< 100$. As we will see later, this is not the case for the ideal MHD model. It is noteworthy that, as for the magnetic field distribution, the distribution of $\beta$ is not evolving much over time except during the build-up phase. 

In Fig.~\ref{fig:correlation_B}, that is similar to Fig.~\ref{fig:correlation_radius} but for the disks vertical magnetic field vs their radius, we clearly see an inverse correlation between the disk size and the vertical magnetic field strength which very close to the $\propto |B_z|^{-4/9}$ (solid black line) predicted by the self-regulated scenario of \citep{2016ApJ...830L...8H}.  Despite being decoupled from the gas at the disk scales, the magnetic field does influence the disk size via the magnetic braking and perhaps via interchange instabilities outside of the disk, i.e., in the envelope.

\subsection{Impact of the magnetic field}
\label{sec:bfield}

\begin{figure*}[h!]
\centering
 \includegraphics[width=
          0.33\textwidth]{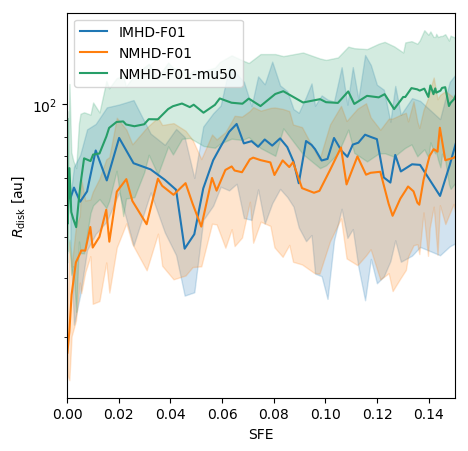}
 \includegraphics[width=
          0.33\textwidth]{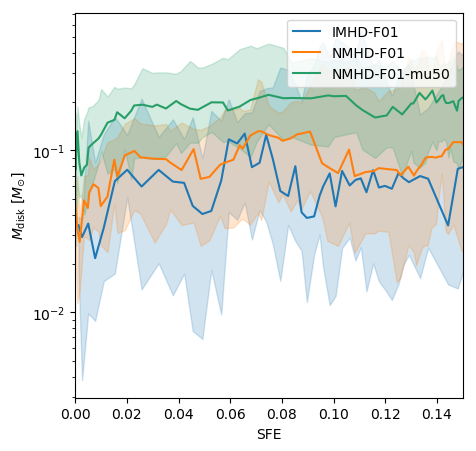}
 \includegraphics[width=
        0.33\textwidth]{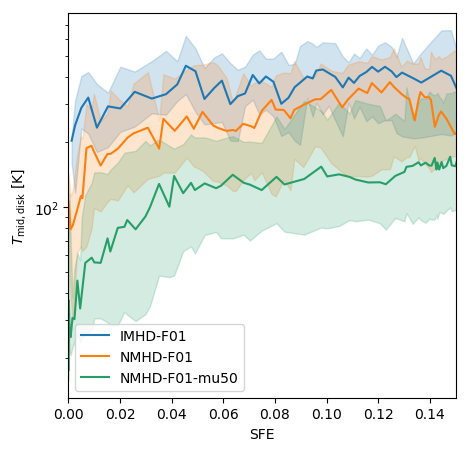}\\
 \includegraphics[width=
          0.33\textwidth]{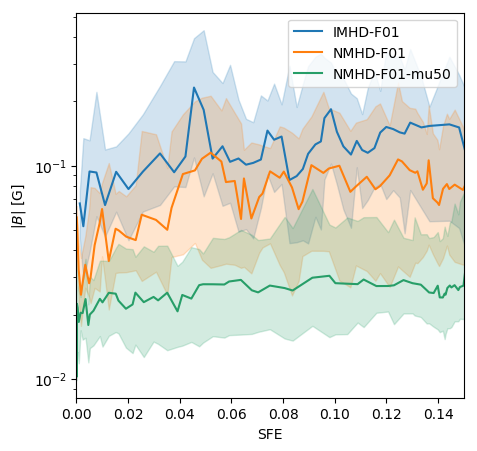}
 \includegraphics[width=
          0.33\textwidth]{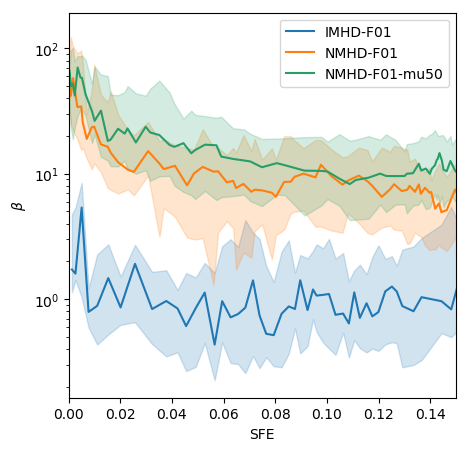}
 \includegraphics[width=
          0.33\textwidth]{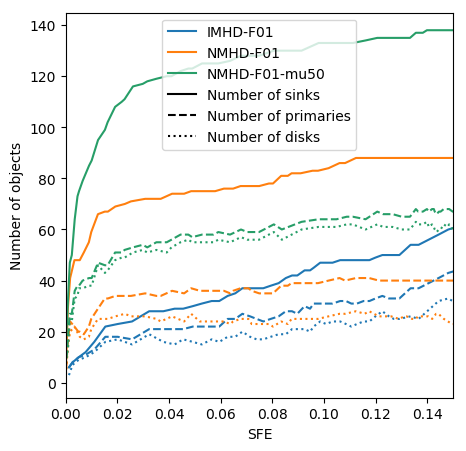}

\caption{Evolution of the disk properties for \nmhd, \imhds and \nmhdw against the SFE. The lines correspond to the median of the distribution and the coloured regions correspond to the area between the first and third quartile of the distribution. From left to right, top to bottom: radius, mass, mid-planet temperature, magnetic field strength, plasma beta and number of objects. } 
              \label{fig:distri_bfield}
\end{figure*}

\begin{figure*}[h!]
\centering
\begin{subfigure}{ 0.33\textwidth}
  \centering
 \includegraphics[width=
          \textwidth]{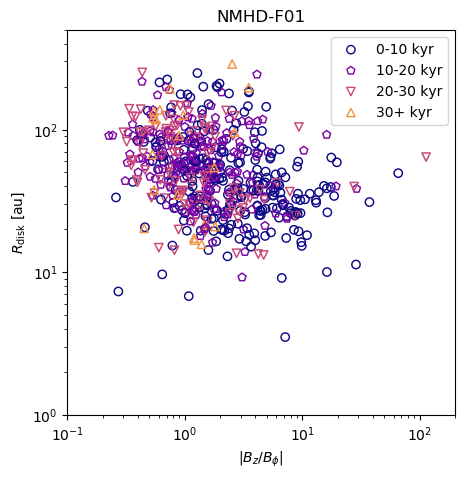}
     \end{subfigure}
         \hfill
     \begin{subfigure}{ 0.33\textwidth}
  \centering
 \includegraphics[width=
  \textwidth]{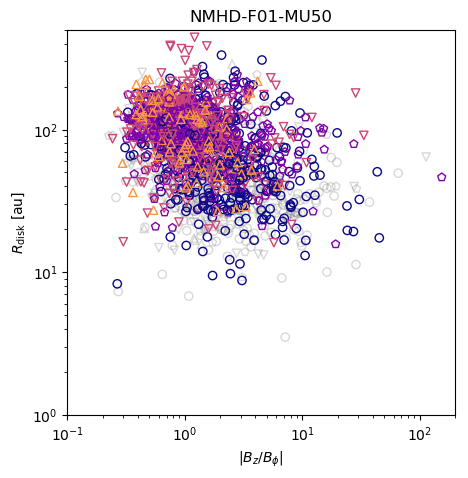}
     \end{subfigure}  
         \hfill 
     \begin{subfigure}{ 0.33\textwidth}
  \centering
 \includegraphics[width=
        \textwidth]{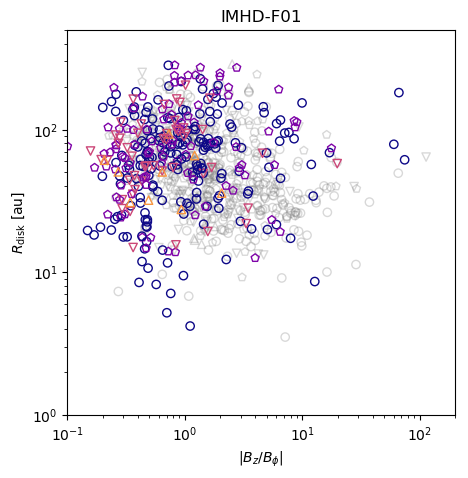}
     \end{subfigure}  
\\
  \begin{subfigure}{ 0.33\textwidth}
  \centering
 \includegraphics[width=
          \textwidth]{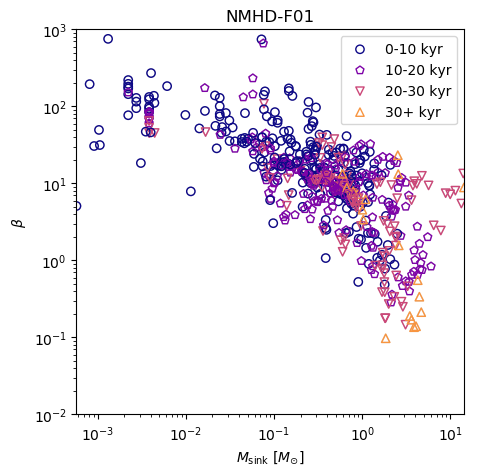}
     \end{subfigure}
         \hfill 
     \begin{subfigure}{ 0.33\textwidth}
  \centering
 \includegraphics[width=\textwidth]{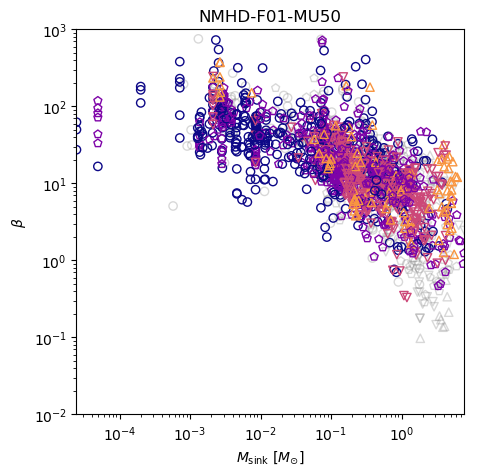}
     \end{subfigure}  
         \hfill
     \begin{subfigure}{ 0.33\textwidth}
  \centering
 \includegraphics[width=
          \textwidth]{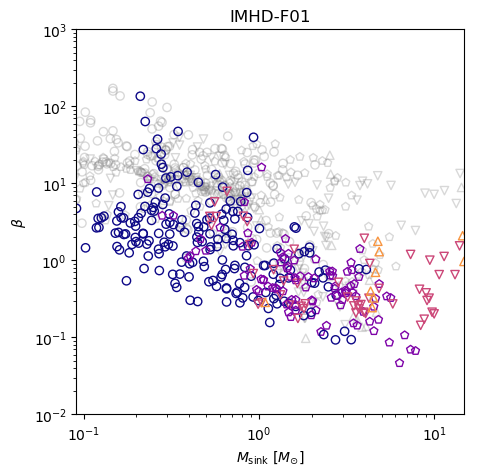}
     \end{subfigure} 
\caption{Correlations between some disk/star properties for runs \nmhd, \nmhdw and \imhd (from left to right). Top: disk size vs ratio of poloidal over toroidal magnetic field . Bottom: Beta plasma vs primary mass. For the \nmhdw and \imhd panels, the information of the \nmhds panel is duplicated with grey markers.  } 
              \label{fig:correl_bfield}
\end{figure*}

\begin{figure}
  \centering
       \includegraphics[width=
         0.45\textwidth]{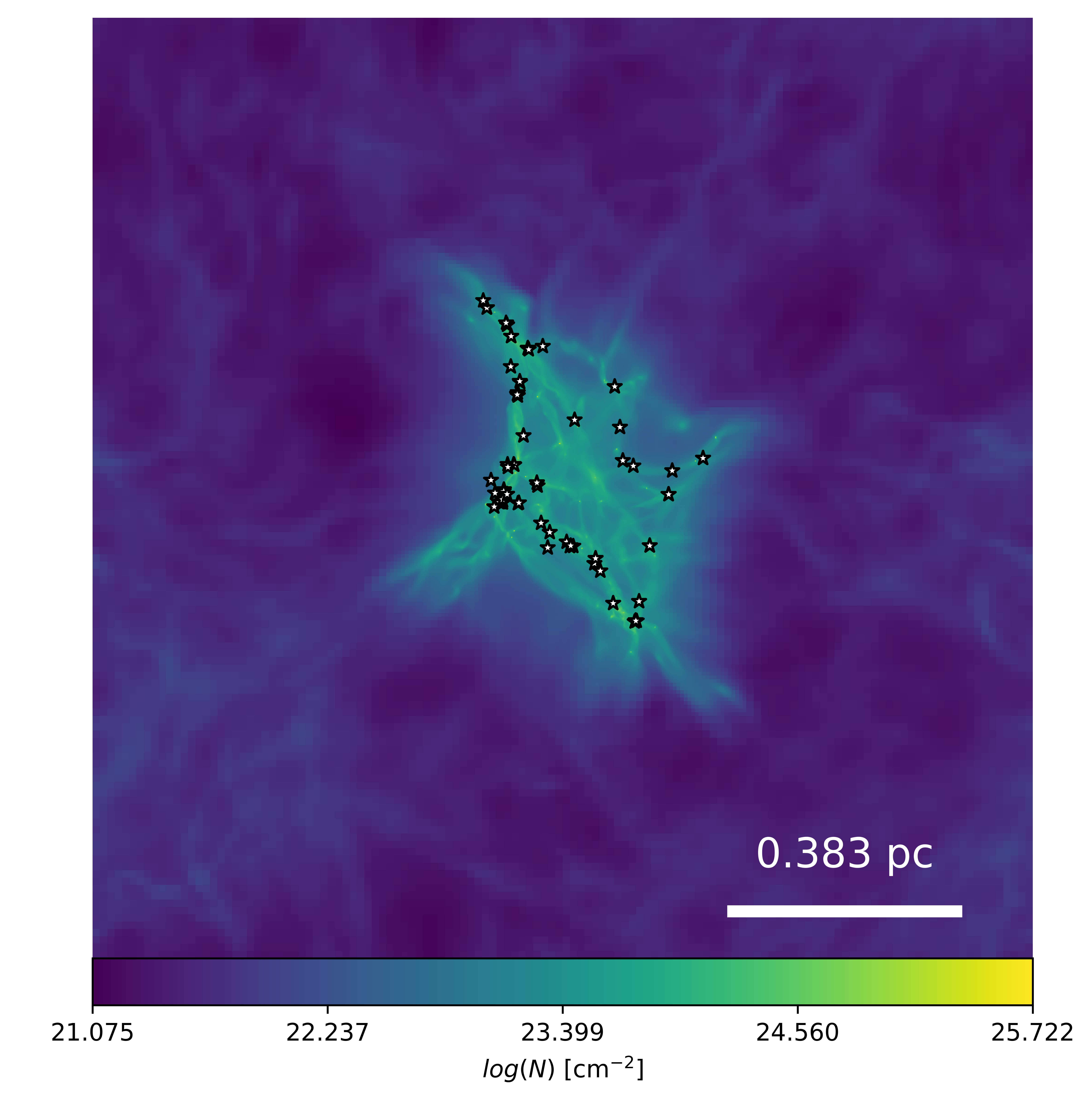}
      \caption{same as panel (c) of Fig.~\ref{fig:column_fid} but for the \nmhdws run. The clump is more fragmented as a result of the lower magnetic pressure support.} 
              \label{fig:nmhdw_coldens}
\end{figure}

We have computed two additional models with a different magnetic field treatment and strength. The first run, \imhd, is the same as \nmhds but evolving the magnetic field with an ideal MHD framework. The second run, \nmhdw, is the same as \nmhds but with a mass-to-flux ratio $\mu =50$, hence an initial magnetic field five times lower than the fiducial value. 

In Fig.~\ref{fig:distri_bfield}, we show the evolution of the disk quantities as a function of the SFE for \nmhd, \imhds and \nmhdw. We show the disk radius (panel a), mass (panel b), mid-plane temperature (panel c), magnetic field  strength (panel d), plasma beta (panel e) and the number of formed objects (stars, primaries and disks, panel f). For panels a to e, the dotted lines represent the median of the distribution and the colored surfaces represent the first and third quartiles of the distribution. Before describing each panel in detail, we note that the disk properties are still relatively steady with time/SFE for all three models (except during the build-up phase).

\paragraph{Fragmentation:} Let us first focus on the number of objects, i.e., panel (f) of Fig.\ref{fig:distri_bfield}. Quite clearly, the magnetic field actively stabilises the cloud against fragmentation in \nmhd. Indeed \nmhdw, that has an initially weaker magnetic field, is able to form more sinks (about $140$, including about $60$ primaries). Consequently, more disks are also formed in this run. Interestingly, almost all the systems are hosting a disk in the model, contrary to \nmhds for which the fraction is closer to $75\%$. This is consistent with what was observed in \cite{Lebreuilly2021} where we have shown that the model without magnetic field, i.e. with an infinite mass-to-flux ratio, forms more stars, more disks and has a higher fraction of disk-hosting stars.  It is worth noting that star formation is happening more homogeneously in the \nmhdws clump that is visibly more fragmented than the one of \nmhd. This indicates that fragmentation is suppressed at a relatively large (larger than the disk) scale in \nmhd. This can be seen in Fig.~\ref{fig:nmhdw_coldens}, that shows the column density and the stars of \nmhdws at SFE=0.15, which can be directly compared with the panel (c) of Fig.~\ref{fig:column_fid}.  Similar observations were shown in \cite{Hennebelleetal2022}. 

If we now focus on the model with ideal MHD, \imhd, we see that it forms even less stars than \nmhd. This is a consequence of the flux-freezing approximation that leads to an overall stronger pressure support of the magnetic field in the ideal MHD calculation. We notice that the fraction of primaries, which is the fraction of stars with no neighbours in a 50 au radius, and also the total number of primaries is however slightly higher that \nmhds at the end of the calculation, because small scale fragmentation is suppressed by the stellar feedback. 

\paragraph{Disk size:} As we can see, whereas there are no strong differences of radii between the ideal MHD and the ambipolar diffusion case \citep[as was also shown in][]{Lebreuilly2021}, a change in the mass-to-flux ratio however has an important impact. The typical disk size is $\sim 30-40~$au in \nmhds and \imhds and about twice larger in the case of a weaker initial magnetic field. In fact, disk sizes are actually slightly larger in \imhd, which we explain later in this section. It is interesting to point out that this difference of a factor of $\sim 2$ is perfectly consistent with the self-regulated scenario of \cite{2016ApJ...830L...8H} from which it is expected that the disk size scales inversely with the square root of the magnetic field strength. The present result is however a generalisation of what is proposed by \cite{2016ApJ...830L...8H}  because, in the present case, the scaling also appears to be valid for the clump scale magnetic field. As pointed out by \cite{2013MNRAS.432.3320S} as well as \cite{2019MNRAS.489.1719W}, the resemblance in terms of disk size between the ideal MHD and ambipolar diffusion models could be a due to turbulence. Indeed it likely diminishes the influence of magnetic braking by generally producing a misalignment between the angular momentum and the magnetic field. However, as we have shown above, the clear correlation between the magnetic field strength and the disk size does indicate that they do play a crucial role in that regard. This is also supported by the fact that we obtain larger disk when considering a weaker magnetic field. We also emphasize that this similarity in the disk radius between the two models is actually misleading because the disk masses of \imhds are lower than those of \nmhds. 

\paragraph{Disk mass:}  The disks of \nmhdws are more massive than those of \nmhds and \imhds by a factor of about 2. This is not surprising since the disk masses are correlated to their radii in our models. Conversely, \imhds disks are generally less massive than those of \nmhd. The efficient magnetic braking at high density, in the absence of ambipolar diffusion, leads to more radial motions even in the disk which allows the star to accrete more material. This is a non-linear effect because more massive stars generate stronger feedback that also reduces fragmentation \citep{2011ApJ...742L...9C}. This aspect is interesting in the light of the similarity of the radii of the disk of \nmhds and \imhd. Essentially, even if their sizes are comparable, ideal MHD disks are typically less dense than the ones with ambipolar diffusion. The similarity in terms of disk size between ideal MHD and MHD with ambipolar diffusion is thus misleading and does not mean that magnetic braking is as efficient in the presence of ambipolar diffusion as it is without.

\paragraph{Disk temperature:} Disks are hotter in \imhds than they are in \nmhd, the disks of \nmhdws being the coolest. In \imhds the fragmentation is suppressed, therefore the stars are able to accrete more material which is further helped by a strong magnetic braking.  Therefore \imhds stars have a stronger accretion luminosity which leads to hotter disks. Conversely, fragmentation is most efficient for  \nmhdws, and on top of that, the disks are larger, which means that a large part of their material is far away from the star and therefore colder.

\paragraph{Disk magnetic field/plasma beta:} Let us focus on the difference of magnetic field properties between the models. As a complement from  panel (d) and (e) of Fig~\ref{fig:distri_bfield}, we show for the three models (left to right) the correlation of the disk size vs the ratio between the vertical and azimuthal magnetic field (hereafter poloidal fraction, top panels) and the plasma beta vs the disk mass (bottom panels) in Fig~\ref{fig:correl_bfield}. 

We see in panel (d) of Fig~\ref{fig:distri_bfield} that, at all SFE, the typical disk magnetic field is stronger in \imhds than it is in \nmhds and unsurprisingly, it is the weakest in \nmhdw. This explains quite clearly why the disks of \nmhdws are the largest in size. As we have shown earlier, the disk size roughly scales as $|B_z|^{-4/9}$ . We see in both \nmhdws and \nmhds
an weak correlation between the disk radius and the poloidal fraction, which is not clear at all for the \imhds model. This could be due to the more efficient winding-up of the magnetic field lines for the more massive disks that is able to generate a significant toroidal field despite the diffusive effect of the ion-neutral friction which was observed in isolated collapse calculation of massive cores \citep{2021A&A...652A..69M,2022A&A...658A..52C}. As we can see (Fig.~\ref{fig:correl_bfield}, panel a) the bottom-right quadrant (small disks, strong poloidal fields) of the plot is dominated by the young 0 to 10 kyr disks that did not have the time to wind up the field, while the top-left quadrant (large disks, weak poloidal fields) is dominated by the older disks. As said earlier, the previously described behaviour is not at all seen in the case of \imhd; moreover, the poloidal fraction is generally lower in this model. This is a key difference that can be explained by the non-negligible impact of the ambipolar diffusion at the envelope scale for \nmhd. Without any diffusive effects, the magnetic field lines are already efficiently wound-up even before disk formation in the collapsing cores and therefore most disks already have a strong toroidal magnetic field at birth time. Because of that, the correlation between the poloidal fraction and the disk size is not observed in the \imhds case.  As explained earlier, the disk sizes of \imhds and \nmhds are not strongly different. In fact, and perhaps counter-intuitively, we find that ideal MHD disks are, on average slightly larger than those of \nmhd. As explained in \cite{Lebreuilly2021}, this is most likely a consequence of the strong toroidal fields, which stabilise the large disks against fragmentation, and that are only formed in ideal MHD.

Quite interestingly, $\beta$ is similar in the disks of \nmhds and \nmhdw, meaning that they reach the same level of magnetisation with respect to the thermal pressure. We clearly see in Fig.\ref{fig:correl_bfield} that the beta plasma decreases with the sink mass for the three models but also that the typical value of beta is way lower for \imhd than it is for the two other models. The majority of \imhds disks are distinctly magnetically dominated rather than thermally, whereas the opposite happens for \nmhds and \nmhdw. Low disk magnetisation is yet another clear consequence of ambipolar diffusion \citep[see][for similar results in low and high mass isolated collapse calculations]{2016A&A...587A..32M,2021A&A...652A..69M,2021A&A...656A..85M,2022A&A...658A..52C}. In ideal MHD disks, the infall actively drags the field lines, which causes a dramatic increase of the magnetic field intensity. In the case of \imhd, because there is no diffusion at the envelope scale, the magnetic field is already strongly enhanced at disk birth. The $\beta$-plasma then does not evolve much over time and stays, on average, close to 1, as can be seen in panel~(e) of Fig.~\ref{fig:distri_bfield}. For the two other (MHD with ambipolar diffusion) models, the magnetic field is, as explained before, mostly vertical at birth. We then observed a slow decrease of $\beta$ with time as a toroidal component is generated by the disk rotation.

 We conclude that although, the disk properties of the ideal MHD and MHD with ambipolar diffusion models present some similarities (radius, temperature), the difference in disk/stellar masses and magnetic field properties between the ideal MHD and MHD with ambipolar diffusion disks leads us to the conclusion that treating the ambipolar diffusion is crucial to better capture the disk formation and evolution. We point out that knowing the initial configuration of the magnetic field is important for the onset and properties of MHD winds \citep[see the review of ][and references therein]{2022arXiv220310068P}. The strength of the magnetic field at the clump scale also appears to be an essential parameter that determines the properties of both the disks and star formation through its impact on the stellar feedback and magnetic pressure that acts against fragmentation.

\subsection{Impact of radiative transfer}
\label{sec:rt_impact}
\begin{figure*}[h!]
\centering
 \includegraphics[width=
          0.33\textwidth]{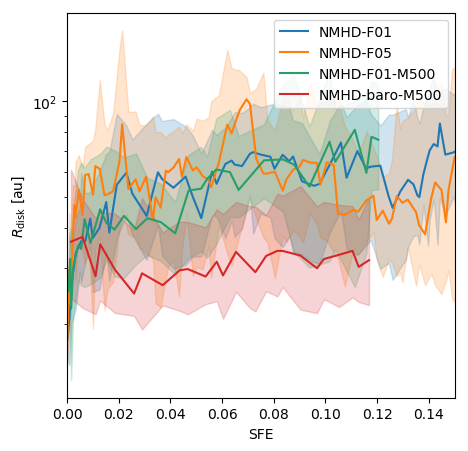}
 \includegraphics[width=
          0.33\textwidth]{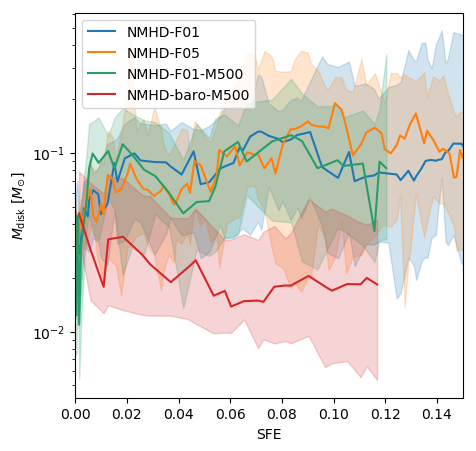}
 \includegraphics[width=
        0.33\textwidth]{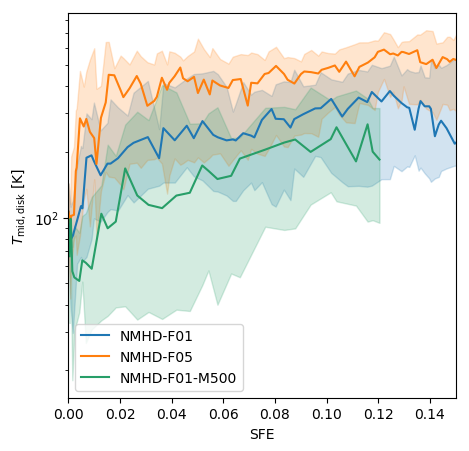} \\
 \includegraphics[width=
          0.33\textwidth]{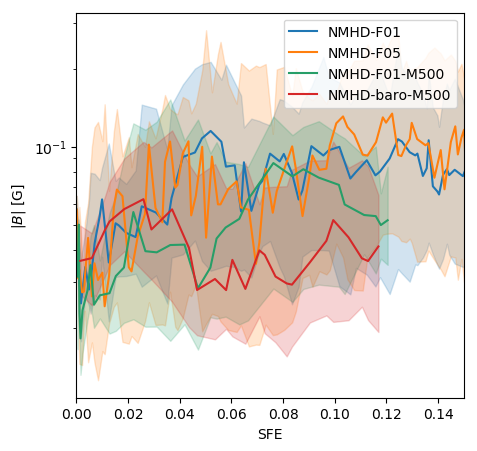}
 \includegraphics[width=
          0.33\textwidth]{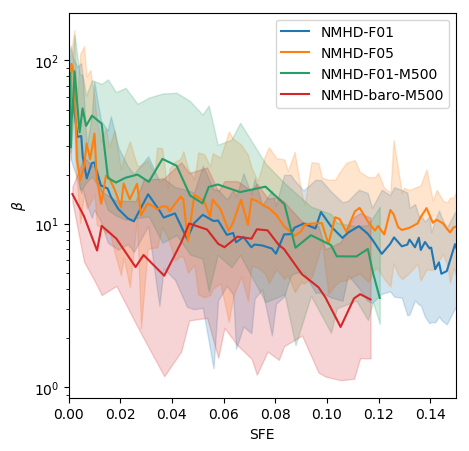}
 \includegraphics[width=
          0.33\textwidth]{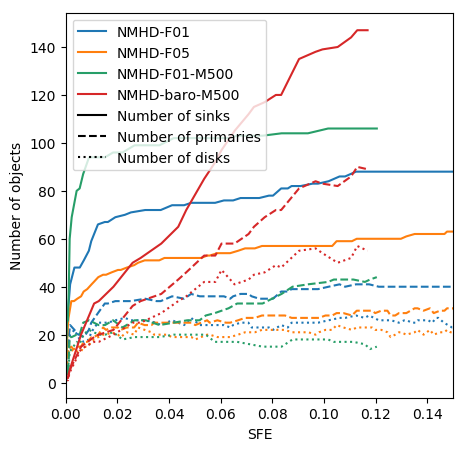}

           \caption{same as Fig.~\ref{fig:distri_bfield} for \nmhd, \nmhdf, \lnmhds and \lnmhdb.}
              \label{fig:distri_rad}
\end{figure*}

\begin{figure}[t!]
  \centering
 \includegraphics[width=
         0.45 \textwidth]{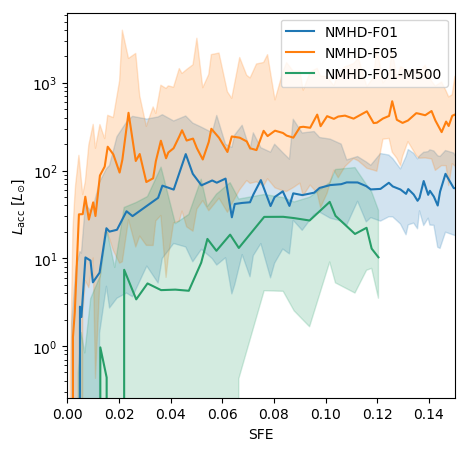}
      \caption{same as Fig.~\ref{fig:distri_rad}
 for the accretion luminosity onto the disk-hosting primaries. The \nmhd, \nmhdfs  and \lnmhds \textbf{runs} are displayed.} 
              \label{fig:accrate}
\end{figure}

\begin{figure*}[h!]
\centering
 \includegraphics[width=
          \textwidth]{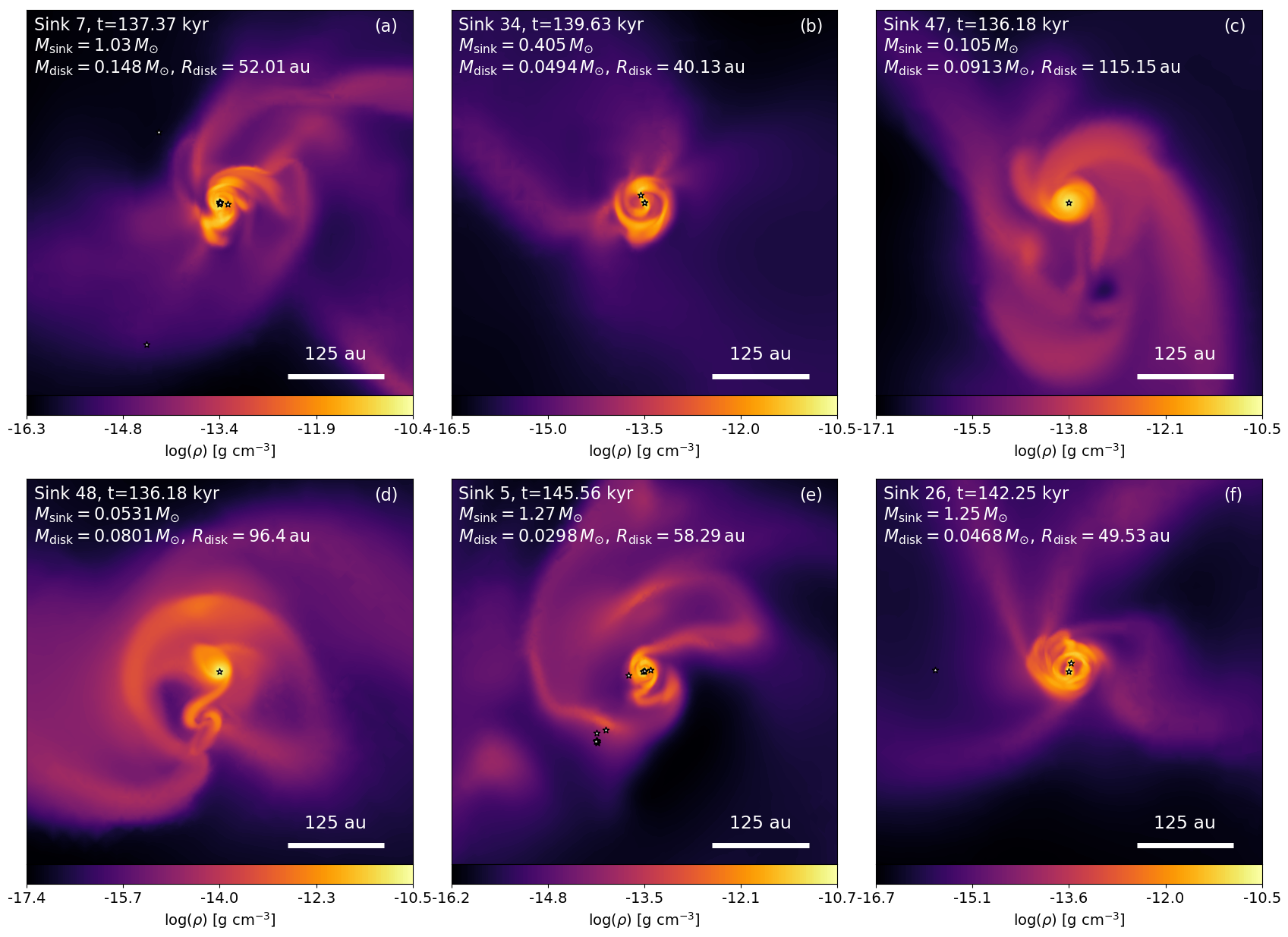}
     \caption{collection of fragmenting disks from run \lnmhdb; mid-plane density slices. In addition to the density, for each disk we display the sink index, the time of the corresponding snapshot, the mass of the sink and of the disk as well as the disk radius. } 
    \label{fig:collec_baro}
\end{figure*}

In this section, we investigate the impact of the modelling of the temperature through the choice of the $f_{\mathrm{acc}}$ parameter and of the RT modeling (FLD, barotropic EOS). We have run the \nmhdfs  model which is similar to \nmhds but with $f_{\mathrm{acc}}=0.5$, therefore we expect a more significant impact of the radiative feedback in this model.  We also computed two additional models to better understand the impact of the RT modelling, \lnmhds and \lnmhdb. To be comparable, both models have been run with $\alpha=0.016$, an initial clump mass of $500 M_{\odot}$, a Mach number of 7 and a mass to flux ratio of 10. However contrary to \lnmhd, that accounted for the RT with FLD and $f_{\mathrm{acc}}=0.1$, \lnmhdbs assumed a barotropic EOS. In a sense, \lnmhdbs gives an idea of what would be the evolution of the clump if no stellar feedback was included. We point out that this parameter choice also allows to probe disk formation in less massive and less dense clumps with respect to \nmhd. In Fig.~\ref{fig:distri_rad}, we show the same information as in Fig.~\ref{fig:distri_bfield}, but for \nmhd, \nmhdf, \lnmhdbs and \lnmhd.

We first focus on the comparison between \nmhds and \nmhdf. If we look at panel (f), we see that fragmentation is even more suppressed with $f_{\mathrm{acc}}=0.5$.
We note that the effect of $f_{\mathrm{acc}}$ is non-linear: suppressing fragmentation leads to more massive stars that are brighter and hotter preventing fragmentation even more. The impact of $f_{\mathrm{acc}}$ is clear when looking at panel (c) of Fig.~\ref{fig:distri_bfield} that shows the mid plane disk temperatures. The typical disk temperature quickly gets higher by a factor $\sim 2$ in \nmhdfs than in \nmhd. The difference in temperature between the two models does not vary much over time which indicates that it is indeed caused by the accretion luminosity. Although $f_{\mathrm{acc}}$ is five times higher in \nmhdfs than in \nmhd, the typical accretion luminosity of the former is almost one order of magnitude higher than in the latter. The additional factor 2 in the accretion luminosity is mostly due to the reduced fragmentation in \nmhdfs that leads to more massive stars.

Despite the difference of disk temperatures, other quantities, such as the disk radius, mass and magnetic field are similar in \nmhdfs and \nmhd. This hints that the thermal support does not strongly affect the formation mechanism of the disks (magnetic braking vs conservation of the angular momentum) and their evolution \citep[quasi Keplerian rotation with a low viscosity as in isolated collapes][]{2020A&A...635A..67H,Lee2021}. We note however that the  disks are typically thicker in \nmhdfs than in \nmhd. This is not surprising at all since the disk scale height is controlled by the thermal support of the disk.

We now focus on the two $500 M_{\odot}$ runs, \lnmhds and \lnmhd. By the end of the calculation, at a SFE = 0.11, the barotropic EOS calculation has formed 140 sinks whereas only 106 have been formed in \lnmhd. This is clearly an effect of the lower thermal support of the \lnmhdb, where the temperature is close to 10 K at low density. It is also interesting to point out that, in the case of \lnmhdb, the maximum star mass is of $\sim 2 M_{\odot}$ whereas it is around $\sim 10 M_{\odot}$ for \lnmhd. Despite being generally less massive (as disk fragmentation is more important in this model), the disks of \lnmhdbs are significantly more unstable due to their low thermal support. This can be seen in Fig.~\ref{fig:collec_baro} that shows examples of fragmenting disks extracted from \lnmhdb. These disks are ubiquitous in the barotropic calculation, but rare for the other models with RT. This is an important point since disk fragmentation could be favourable for planet formation, either through streaming instability in the pressure bumps or through gravitational instability.

 It is also worth mentioning that \lnmhds and \lnmhdbs have a lower initial density that \nmhd. As a consequence, the sink accretion rates are generally lower and so is the accretion luminosity (see Sect.~\ref{sec:acclum}). If this has no consequences for \lnmhdb, which is computed with the barotropic EOS, the disks of \lnmhds are impacted by this lower accretion rate and are colder than the ones of \nmhd. This confirms that controlling the accretion rate is crucial to predict the disk temperature and hence its fragmentation. As for the comparison between \nmhds  and \nmhdf, it is interesting to point out that, temperature aside, the disk of \nmhds and \lnmhds are however quite similar. In addition for showing that the temperature is apparently not a controlling factor for the disk size, mass and magnetisation (unless the disk are indeed very cold), this seems to indicate that the clump mass, or rather its initial density, is not either.

 We conclude that a precise modeling of the RT including the impact of the accretion luminosity seems to be crucial to constrain the disks temperatures and the clump fragmentation. In addition, unless very cold disks are somehow relevant (if the barotropic EOS latter proves to be good approximation), the choice of accretion luminosity efficiency does not impact much the size and mass of the disks.
\section{Discussion}

\subsection{Comparison with observations}
\label{sec:obs}
\begin{figure}[h!]
\centering
 \includegraphics[width=
          0.45\textwidth]{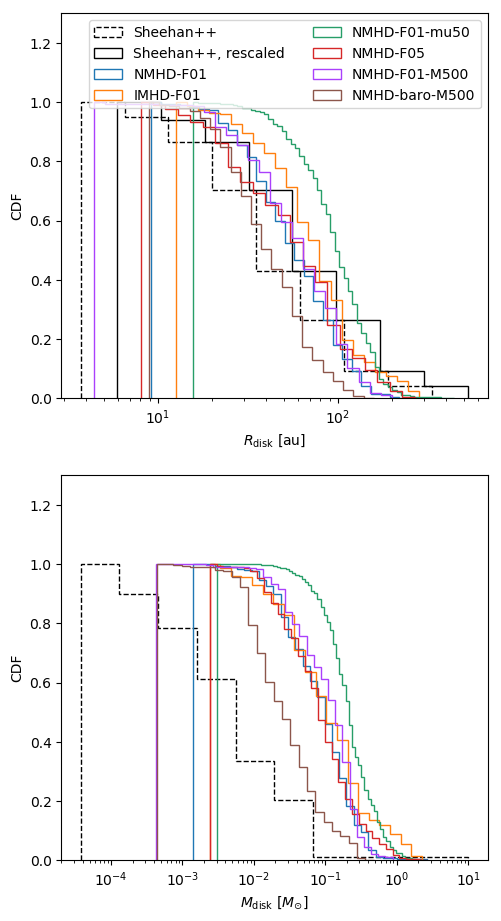}
     \caption{comparison of the disk distributions of all our models (lines) with the observed ones in a sample of protostars in the Orion molecular cloud. A good agreement is found for the disk sizes but an important tension is observed when it comes to the disk mass. The observations uncertainties are not displayed for readability. } 
    \label{fig:obs_sim}
\end{figure}

One of the primary goals of the \textit{Synthetic Populations of Protoplanetary Disks} project is to provide models to compare simulations and observations of Class 0 disks statistically.

We present a first tentative comparison of the disk populations with observed ones. For that comparison, we use survey of disks in the Orion cloud \citep[VANDAM survey,][]{2022ApJ...929...76S}.  In Fig.~\ref{fig:obs_sim}, we show the cumulative distribution of the disk radius and mass of the populations extracted from all our models (coloured lines) compared with the observed ones (black lines). For the observations, we display the brute data of the survey with dashed lines and a version, re-scaled them by a factor 1/0.63, with plain lines. This re-scaling is done because the truncation radius as used in the \cite[VANDAM survey,][]{2022ApJ...929...76S} could be a better estimate of the radius containing 63$\%$ of the disk mass while our estimate better approximates the total disk radius. This crude re-scaling gives a good idea of what the total radius would be in this survey. In our populations, the disks are sampled every 1~kyr to mimic a diversity in evolutionary stages and to enhance the statistics in our clumps. We  only selected the disks after the build-up phase, i.e., after 10 kyr, to make sure that the distributions are in their steady phase, more likely to be observed. We also kept the populations of the different clumps separated to see the impact of the initial clump scale properties/physical assumptions on the disks.

In terms of disk radii, there is a moderately good agreement between our models and the observations. We note that the agreement is best for the models (with ambipolar diffusion) with a stronger initial $\mu=10$ magnetic field. Conversely, the \nmhdws model, that has an initial $\mu=50$, consistently produces discs larger than those observed. As was noted by \cite{2018MNRAS.475.5618B}, the radius that contains 63$\%$ of the mass could be a better measure of the radius in comparison with observations when a truncated power laws is assumed to fit the disks. Considering that this value is more likely the one measured in the observations actually makes the agreement with the $\mu=10$ models even better. We however point out that there is only a factor $\sim 2$ difference between the disk radius across all the populations. In addition it is important to stress a very important point. Disk observations are actually sensible to the continuum flux of the dust and not the gas mass. The conversion from this flux to mass is not at all trivial, as we discuss below, therefore it is not clear at all which of the $r_{63}$ or $r_{\mathrm{disk}}$, if any, is actually better probed by observations. This issue might be partly solved in \cite{2022ApJ...929...76S} though, as they that performed a careful radiative transfer modelling to fit the observed disks. We will address this issue in a upcoming work, were we post-process our models to produce synthetic observations. This will allow us to compare our models to real observations extracting the disks with the exact same methods.

We also recall that we found good agreement with disks radii from the CALYPSO survey \citep{2019A&A...621A..76M} in the previous models of \cite{Lebreuilly2021}. At that time, the magnetically regulated models were also in better agreement with the observations. This is also consistent with the observational evidence that show that only magnetically regulated models of the evolution of solar-type protostellar cores, with mass-to-fluxes ratio $\mu < 6$ could reproduce the disk properties of the B335 protostar \citep{2018MNRAS.477.2760M}. 

There is a more significant tension between our models and the observations when it comes to the  disk mass. We generally report more massive disks than those obtained in \cite{2022ApJ...929...76S}. We point out that similar tensions between models and observations were previously reported by \cite[][Fig. 25]{2018MNRAS.475.5618B} and \cite[][Fig. 9]{2020ApJ...890..130T} for the disk masses. As explained earlier, the masses obtained with our models are in line with those reported by \cite{2018MNRAS.475.5618B} despite the significant differences in numerical methods. They indeed report, as we do, typical disk masses between $<0.01$ and $1 M_{\odot}$. We point out that the disks of \cite{2022ApJ...929...76S}, from the Orion cloud actually have lower masses than those of Perseus \citep{2020A&A...640A..19T} and Taurus \citep{2017ApJ...851...45S}. Considering that the disk mass depends on the environment might partly solve the problem, as pointed out by \cite{2021MNRAS.508.5279E} (their Fig. 9). At this stage it is worth mentioning that the disk masses are poorly constrained both observationally and theoretically. Disk masses are likely controlled by the relatively unknown inner boundary condition i.e., the star-disk interaction (see Sect.~\ref{sec:caveats}). On the observational side, the arbitrary choice of dust model and size distribution can lead to potentially large errors in the conversion between the flux from thermal dust emission measure at mm wavelengths, and the disks mass. This issue, discussed in the recent review by \cite{2022arXiv220913765T}, could be significant as the dust optical properties have been obsered to be very different in protostellar environments than in the diffuse ISM \citep[e.g.,][]{2019arXiv191004652G}. In addition, the computation of the gas mass from the dust relies on a conversion that assumes a constant gas-to-dust ratio, which is usually chosen to be 100. This hypothesis could be wrong in both ways: dust-rich disks \citep[that can form if the dust decouples from the gas during the collapse][]{2020A&A...641A.112L} were the gas-to-dust ratio could be lower than 100 and the disk mass would be overestimated, and dust-poor disks, for example if some of the dust has already been converted into planetesimals, the gas-to-dust ratio would be larger than 100 and the disk mass would be underestimated. Unfortunately, molecular tracers might not lead to better estimates for the same reason as they also rely on a conversion factor to get the H2 disk mass. The gas kinematics could provide us with dynamical estimates of the disk mass assuming that the protostar mass is known \citep{2021ApJ...907L..10R,2023MNRAS.518.4481L}. Although this methods are challenging, they might be our best hope for a precise inference from the observational point of view.

 It is also important to point out that most of the observed protostellar disks, including Class 0 disks, are older than 50 kyr, while our older disks are 'only' about 40 kyr by the end of our simulations. Their properties could, in principle evolve quite a lot through the Class 0 leading to an apparent disagreement between the mass estimates from models and observation that is, in fact, only due to an age difference. However, as we explained earlier, the disks properties are not significantly varying in our models for disks older than $\sim 10~$kyr and, as was shown in isolated calculations \citep{2020A&A...635A..67H}, this relatively steady state probably lasts for at least 100 kyr. This supports us into thinking that there might indeed me a fundamental disagreement between models and observations for the disk masses.

Finally, and quite surprisingly, the \lnmhdbs model seems to be the model that actually fit better the observations for both the mass and the radius. We however stress that this likely is a coincidence. Barotropic models are indeed not supported by the recent e-disk survey \citep{2023ApJ...951....8O} that have shown that young disks are quite hot and do not show obvious sub-structures except in rare and evolved case. This survey is thus is in much better agreement with our RT models.

\subsection{On the luminosity problem}
\label{sec:acclum}

\begin{figure*}[h!]
\centering
\begin{subfigure}{0.45\textwidth}
  \centering
 \includegraphics[width=
          \textwidth]{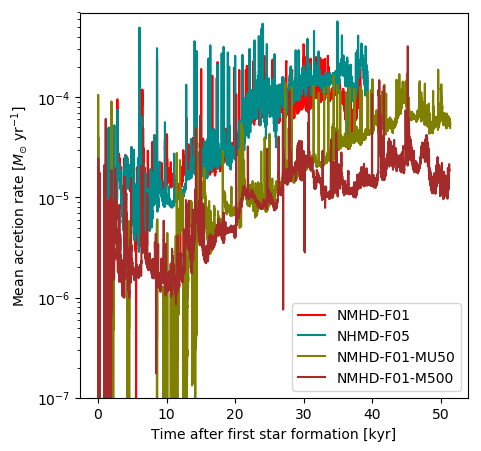}
     \end{subfigure}
     \hfill
 \begin{subfigure}{0.45\textwidth}
  \centering
 \includegraphics[width=
          \textwidth]{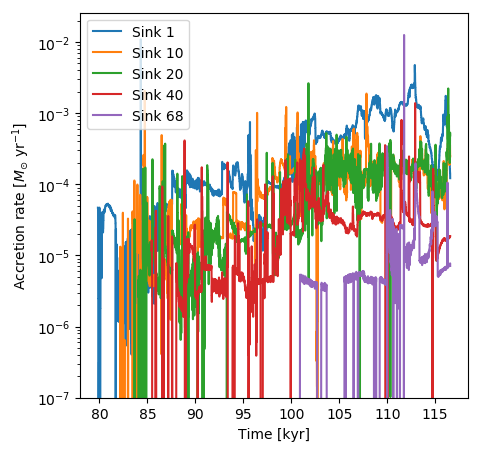}
     \end{subfigure}
     \caption{Evolution of the stellar accretion rate as a function of time. (Left) mean accretion rate for \nmhds,  \nmhdf, \nmhdws and \lnmhds  as a function of the time since the first star has formed. (Right) Same but for the individual accretion rates of a collection of sinks of \nmhd.  } 
    \label{fig:accrates}
\end{figure*}

The luminosity problem \citep{1990AJ.....99..869K} is a long-standing issue of star formation. Observed YSO luminosities are below the values expected from  steady-state protostellar accretion. This hints that either some of the accretion luminosity is not fully radiated away or the accretion is highly variable during protostar formation. Both of these issues can in principle be taken into account in the model through the efficiency factor $f_{\mathrm{acc}}$ which is a sub-grid modeling for the conversion of accretion luminosity into radiation in our calculations. 

The typical accretion luminosities of our models with $f_{\mathrm{acc}}=0.1$ are typically a few $10~L_{\odot}$, while they are about one order of magnitude higher in the case of \nmhdfs that has $f_{\mathrm{acc}}=0.5$. Conversely, YSO observations seem to indicate lower luminosities with typical values of the order of a few $L_{\odot}$ during  the Class 0 stage \citep{2011A&A...535A..77M,2017ApJ...840...69F}.

As explained earlier, the accretion luminosity might not have a significant impact when it comes to the disk masses, sizes and the properties of the magnetic field.  However, constraining the accretion luminosity is of paramount importance for the disk temperature and our understanding of planet formation in those disks. First, the thermal support brought by stellar irradiation can act against the formation of structures in the disks \citep[][see also the comparison between \lnmhdbs and \lnmhd]{2011MNRAS.418.1356R}. In addition, the position of the snow lines depends on the temperature profile of the disk, which is important since planetesimal formation is expected to be most efficient at their vicinity where the gas and dust properties of the disks abruptly change \citep[see][and references therein]{Drazkowska2022}. The position of the snowlines also determines the composition of the material available to form planets (gas and solids) at different locations in the disk. For instance, the location of the H$_2$O snowline is crucial to understanding in which conditions rocky planets form and how water is delivered to them. At the same time, in a relatively hot disk, some volatile species would never condense. Connected to that, a growing body of literature is studying how to link chemistry in disks to planet formation and to the composition of the exoplanets that we observe today \citep[see for instance][]{2021PhR...893....1O,2021ApJ...909...40T,Pacetti2022}. All of these works will considerably benefit from better constraints on $f_{\mathrm{acc}}$.

At this stage it is important to emphasize that the real and effective value of $f_{\mathrm{acc}}$, if constant at all, could, in principle be even lower than $0.1$. In this case, there must either be an efficient mechanism to convert the gravitational energy into something else than radiative energy (for example magnetic energy or internal convective energy) or most of the radiation should be lost through the outflow cavity. As we will show in Sect.~\ref{sec:caveats}, this issue could be tackled by models that resolve the star-disk connection up to the small scales, i.e. the stellar radii.

As previously mentioned, accretion in strong bursts could be invoked to solve the accretion luminosity problem \citep{2011ApJ...736...53O,2012ApJ...747...52D,2022MNRAS.517.4795M,2023MNRAS.518..791E}. In fact, large variations of luminosity over the timescale of years have been reported for some protostars \citep[for example for B335][]{2023ApJ...943...90E}. This is important because with a strong and steady accretion rate, the disk would be consistently kept warm, whereas short burst of accretion would not have long-term consequences on the disk temperature as the cooling timescale by the dust is very short \citep{2020ApJ...904..194H} compared with the free-fall timescale. Figure~\ref{fig:accrates} shows the mean accretion rate of the sinks as a function of time for \nmhd,  \nmhdf, \nmhdws and \lnmhds (left) as well as the individual accretion rate of some sinks of \nmhds (right). For all the models, there is a clear variability of the accretion rate over time. Accretion indeed occurs in short (but frequent) bursts of a few hundred of years. During those bursts the accretion rate rises to around $10^{-3} M_{\odot} ~\mathrm{yr}^{-1}$, occasionally reaching up to $10^{-2} M_{\odot} ~\mathrm{yr}^{-1}$. However, the average accretion rates are generally high in \nmhds and \nmhdfs in spite of these strong bursts; for these models, we observed typical average values around $10^{-5}-10^{-4} M_{\odot} ~\mathrm{yr}^{-1}$. This explains why, despite having $f_{\mathrm{acc}}=0.1$, the accretion luminosity of our protostars is above typically observed values. Interestingly, the accretion rate is significantly lower in the case of \lnmhds where it is typically around $10^{-6}-10^{-5} M_{\odot} ~\mathrm{yr}^{-1}$. This clump being less dense than the fiducial run, it is also significantly colder (because it cools faster) and more efficiently fragmenting. Similarly, the weaker magnetic field of \nmhdws leads to more fragmentation and lower accretion rates. We point out that episodic accretion is possibly, if not likely, not fully resolved in out models both in terms of space and time. With shorter and stronger bursts, we could expect a lower typical average accretion rate while still accreting enough stars to build the protostar. 

 It is not clear at this stage that there is a luminosity problem in these models. The typical accretion rate indeed depends on the initial conditions of the clump and, as a result, the accretion luminosity can vary by a few orders of magnitude between the various models. We suspect that the various clumps that we present in this study are representative of different star forming regions. With that in mind, it is worth pointing out that \nmhd/\nmhdfs are probably more representative of massive star forming regions whereas \lnmhds is probably better reproducing less compact nearby clumps. We indeed have a top-heavy IMF in the case of \nmhd, contrary to the other runs, and as also shown in \cite{Hennebelleetal2022}. An in-depth exploration of a larger parameter spaces in such simulations and of distant clumps in observations \citep[e.g,][]{2017MNRAS.471..100E,2022A&A...662A...8M} would be required to confirm that hypothesis.

 \subsection{Caveats}
 \label{sec:caveats}

 \subsubsection{The star-disk connection}

A maximal resolution of $\sim1$~au in the disks is already considerable with  simulation boxes of $\sim 1.5$~pc. Unfortunately, this minimum cell size still does not allow to resolve the disk-star connection. To achieve that goal, one should be able to resolve the stellar radii, i.e., to increase the resolution by at least two if not three orders of magnitude. This is, of course, still impossible in the context of disk formation simulations. It is even more challenging in our case because we need to integrate the disks for a few tens of thousand of years to get a steady disk population. Because of that difficulty, we have to rely on the widely used sink particles \citep{1995MNRAS.277..362B,2010ApJ...713..269F,2014MNRAS.445.4015B} as they allow to integrate the model for a much longer time with a somewhat realistic inner boundary condition for the disk. It is important to recall that, unfortunately, the choice of the sink parameters (accretion threshold $n_{\mathrm{thre}}$, sink radii and $f_{\mathrm{acc}}$) does affect the calculation. \cite{2020A&A...635A..67H} showed that the mass of the disk was particularly affected by the choice of $n_{\mathrm{thre}}$ because the density at the center of the disk is quickly adjusting to this threshold. It is however worth mentioning that the choice of this parameter is not completely arbitrary, we have indeed chosen $n_{\mathrm{thre}}$ based on the $\alpha$-disk estimate of \cite{2020A&A...635A..67H}. Fortunately, as they have shown, the disk radius is much less affected by the choice of $n_{\mathrm{thre}}$, probably because it is rather controlled by the magnetic field \citep{2016ApJ...830L...8H}.

In addition, we have shown in Sect.~\ref{sec:rt_impact} that the impact of $f_{\mathrm{acc}}$ on the disk size, mass and magnetic field is probably limited, provided that the accretion luminosity is not negligible (hence in all our models except for \lnmhdb). At the same time, $f_{\mathrm{acc}}$ does affect both the disk temperature and fragmentation. 

To better constrain these two essential parameters, we strongly encourage studies dedicated to understanding the star-disk connection through the modelling of the stellar scales \citep{2018A&A...615A...5V,2020A&A...638A..86B,2023arXiv231001516A} while integrating the models in time as much as possible \citep{2020A&A...635A..67H}. This would provide the community with the necessary sub-grid modelling (star-disk connection) to constrain better the initial mass and temperature of protostellar disks.

 \subsubsection{Dust and planets}
 
 Recent studies indicate that dust evolution is not negligible during the protostellar collapse at disk-like densities \citep{2020A&A...643A..17G,2020A&A...637A...5E,2021ApJ...920L..35T,Bate2022,Kawasaki2022,Lebreuilly2023,Marchand2023}. This of course has important implications for the coupling between the gas and the magnetic field by means of the magnetic resistivities, and probably on the RT because the opacities are dominated by the dust contribution.

 In our collapse calculations, the dust size distribution is, as often, assumed to be a constant Mathis, Rumpl, Nordsieck (MRN) distribution \citep{1977ApJ...217..425M} and is only used to compute the resitvities and the opacity. Its evolution should, in principle, be taken into account self-consistently. 
 This is, unfortunately, still very challenging in 3D simulations. In particular, in the context of large star forming clumps the memory cost of simulating multiple dust grains sizes would be extremely high. Fortunately, recent developments of new methods based on the assumption of growth purely by turbulence  \citep{2021A&A...649A..50M,Marchand2023} or accurate dust growth solvers that require only a few dust bins \citep{2021MNRAS.501.4298L} are opening the way to account for dust in future 3D simulations. 

It is also worth mentioning that sufficiently large grains can, in principle, dynamically decouple from the gas \citep{2017MNRAS.465.1089B,2019A&A...626A..96L,2020A&A...641A.112L,2022MNRAS.515.6073K}. This phenomenon can lead to local enhancement/depletion of the dust material, which would affect the dynamical back-reaction of the grains on the gas. In disks, this mechanism could also self-trigger the formation of sub-structures only in the dust \citep[e.g.,][]{2015MNRAS.453L..73D,2019MNRAS.483.4114C,2020A&A...639A..95R}. Fully including the dust dynamics would also require to take into account the dust growth. This is clearly beyond the scope of this investigation.

Last but not least, as our models do not follow dust evolution in the disks, we are thus unable to make predictions for planet formation and its implication. Again, this is way beyond the scope of the present study. It is however important to keep in mind that fully formed planets or even planetary embryos could probably perturb the disk evolution and trigger the formation of sub-structures \citep[e.g.,][]{2016MNRAS.459L...1D,2017ApJ...850..201B}. To the best of our knowledge, planet population synthesis methods \citep[see the review by][]{2014prpl.conf..691B} are only used in 1D disks; however, their employment could be a way to tackle the problem in future works.

\subsection{Hall effect and Ohmic dissipation}

Ambipolar diffusion is not the only non-ideal MHD effect with a potential effect on disk formation processes. If the Ohmic dissipation is probably only dominant at very high densities \citep{2016A&A...592A..18M} and thus can be more safely neglected, the Hall resistivity might be comparable, if not larger than the ambipolar resistivity in a significant density range in protostellar envelopes and perhaps also in the disks \citep{2021MNRAS.501.5873W}. For numerical reasons, we could not run our simulations with the Hall effect, but it is useful to recall its expected impact from our knowledge of isolated collapse calculations. 

The Hall effect was indeed investigated in details by several groups over the past decade \citep[e.g.][]{2011ApJ...738..180L,2015ApJ...810L..26T,2016MNRAS.457.1037W,Hall-pierre,2017PASJ...69...95T,2019MNRAS.486.2587W,2019A&A...631A..66M,2019MNRAS.489.1719W,2020MNRAS.492.3375Z,2021MNRAS.505.5142Z,2021MNRAS.507.2354W,2021ApJ...922...36L}. These works typically have found that the Hall effect could either enhance or decrease the rotation of the cloud (and the disk) depending of the initial relative orientation between the magnetic field and the angular momentum. Consequently, the Hall effect is expected to produce a bi-modality in disk properties. In the case where the Hall effect would enhance the rotation of the disk, therefore decreasing the effect of the magnetic braking, counter-rotating envelopes can typically be observed. It is worth noting that the Hall effect could also play a role in the fragmentation of the disks, when it accelerates rotation, as was shown by \cite{2019A&A...631A..66M}.

It is not clear yet whether these effects are to be expected to play a strong role in the birth of disk populations obtained from a star forming clumps because of the dispersion in the relative orientation of the magnetic field and the angular momentum.  So far, only \cite{2019MNRAS.489.1719W} investigated disk forming in star forming clumps with all three non-ideal MHD effects and have found no strong impact in the disk size. We point out that we also find a similar trend in our pure ambipolar diffusion case, although, as we pointed out, ideal MHD disk do have lower mass than the ones obtained with ambipolar diffusion. In addition, our ideal MHD disks do not have the same magnetic properties as the ones with ambipolar diffusion. In any case, simulations of massive star-forming clumps including the Hall effect resolving the disks scales would be very valuable for the community and should surely be performed in the future years.

 \section{Conclusion}

In this work, we have explored the formation of protostellar disk populations in massive protostellar clumps with various assumptions and initial conditions. We now recall the main findings and conclusions of this work.

\begin{itemize}
    \item Disk populations are ubiquitous in these simulations. A disk is found around $70$ to $90\%$ of the stellar systems depending on the clump initial conditions.
    
    \item Disks are born with a variety of sizes, masses, structures and nearby environments reflecting their individual history in a highly interacting gravo-turbulent star forming clump.
    \item We commonly find compact disks (in the presence of a strong magnetic field), non-axisymmetric envelopes (accretion streamers), sub-structures (spirals, arcs), magnetised flows (interchange instabilities), flybys and peculiar structures such as disks formed in a single column. However, we find no ring structures or protostellar outflows/jets. 
    \item Accretion luminosity is the dominant source of heating in the disks and it is controlling their temperature. The strength of the accretion luminosity depends on the clump properties. Clumps that fragment more efficiently also have lower accretion rates/luminosity resulting in colder disks.
    \item The strength of the magnetic field at the clump scale is found to be a controlling factor for the disk size and the clump scale fragmentation. A stronger magnetic field leads to typically smaller disks and a reduced fragmentation. Because of that, the disks in clumps with a stronger initial magnetic field are also hotter. 
    \item The accretion luminosity does not seem to be a controlling factor for the other disk properties  (size, mass and magnetic field). However, we point out that these properties could change if the disks were much colder than expected.
    \item The disk sizes obtained from our models are in relatively good agreement with the observed protostellar disk sizes from millimeter surveys. Depending on the way the disk radius is measured, either the case $\mu=10$ or the case $\mu=50$ better fit the models. However, there is tension with some surveys concerning the masses. Future post-processing of the models with radiative transfer tools should clarify the comparison between models and observations.
\item Some well known properties of isolated collapse calculations still hold in the context of large scale models. We confirm the important role of the magnetic field in shaping the disk masses and sizes and its combined importance with the radiative transfer in controlling their temperature and fragmentation. In addition, we show that when we account for ambipolar diffusion disks are weakly magnetized while ideal MHD disks are not. Similarly to high mass star collapse calculations, we find that more massive disk generate stronger toroidal magnetic fields. Finally, we find that disks obtained in barotropic calculations are more easily fragmenting than those of RT calculations. This confirms the interest of the isolated collapse approach to model protostellar disk formation despite its incapacity to provide us with the statistics of the disk populations.  

\end{itemize}

In this work, we have shown how diverse the populations of protoplanetary disks can be at early stages and how they depend on their large scale environment (magnetic field, radiation, cloud mass) as well as on the physical effects included (magnetic field with and without ambipolar diffusion, radiative transfer). We strongly encourage future works further exploring the influence of more clump properties (turbulence, size, shape) as well as dedicated studies comparing such models with real observed data with synthetic observations produced with elaborated radiative transfer codes. 

 \begin{acknowledgements}
We thank the referee for providing a very constructive report that helped us a lot to improve the quality of our paper. This research has received funding from the European Research Council synergy grant ECOGAL (Grant : 855130). We acknowledge PRACE for awarding us access to the JUWELS super-computer.
GR acknowledges support from the Netherlands Organisation for Scientific Research (NWO, program number 016.Veni.192.233). Funded by the European Union (ERC, DiscEvol, project number 101039651). Views and opinions expressed are however those of the author(s) only and do not necessarily reflect those of the European Union or the European Research Council Executive Agency. Neither the European Union nor the granting authority can be held responsible for them.
RSK furthermore acknowledges financial support from the German Science Foundation (DFG) via the collaborative research center (SFB 881, Project-ID 138713538) “The Milky Way System” (subprojects A1, B1, B2 and B8), from the Heidelberg Cluster of Excellence “STRUCTURES” in the framework of Germany’s Excellence Strategy (grant EXC-2181/1, Project-ID 390900948), and from the German Ministry for Economic Affairs and Climate Action in project “MAINN” (funding ID 50OO2206). RSK also thanks for computing resources provided by the Ministry of Science, Research and the Arts (MWK) of the State of Baden-W\"{u}rttemberg through bwHPC and DFG through grant INST 35/1134-1 FUGG and for data storage at SDS@hd through grant INST 35/1314-1 FUGG.
MG acknowledges financial support from the French Agence Nationale de la Recherche (ANR) through the project COSMHIC (ANR-20-CE31-0009).
\end{acknowledgements}
%-------------------------------------------------------------------
\bibliographystyle{aa}
\bibliography{main}

\end{document}